\documentclass[reprint, aip, groupedaddress, amsmath, amssymb, prl]{revtex4-1}
\usepackage{graphicx}
\usepackage{dcolumn}
\usepackage{bm}
\usepackage[normalem]{ulem}
\usepackage{amsmath}
\usepackage[english]{babel}
\usepackage{braket}
\usepackage{units}
\usepackage[T1]{fontenc}
\usepackage[dvipsnames]{xcolor}

\newcommand{\change}[1]{\textcolor{black}{#1}}

\newcommand{\Hamiltonian}{\hat{\mathcal{H}}}
\newcommand{\linmode}[2]{{\hat{#1}_{#2}^\dagger \hat{#1}_{#2}}}

\newcommand{\what}[0]{\hat{\textrm{w}}}
\newcommand{\chat}[0]{\hat{c}}
\newcommand{\sg}{\text{\textit{i}SWAP}}
\newcommand{\rtsg}{\sqrt{\text{\textit{i}SWAP}}}

\begin{document}

\preprint{APS/123-QED}

\title[]{Realizing all-to-all couplings among detachable quantum modules using a microwave quantum state router}

\makeatletter
\def\@fnsymbol#1{\ensuremath{\ifcase#1\or \dagger\or \ddagger\or
   \mathsection\or \mathparagraph\or \|\or **\or \dagger\dagger
   \or \ddagger\ddagger \else\@ctrerr\fi}}
\makeatother

\author{Chao Zhou$^{1\dagger}$, Pinlei Lu$^{1\dagger}$, Matthieu Praquin$^2$, Tzu-Chiao Chien$^1$, Ryan Kaufman$^1$, Xi Cao$^1$, Mingkang Xia$^1$, Roger S.~K.~Mong$^1$, Wolfgang Pfaff$^3$, David Pekker$^1$, Michael Hatridge$^1$}

\affiliation{$^1$Department of Physics and Astronomy, University of Pittsburgh, Pittsburgh, PA, USA \\ 
$^2$Département de Physique, École Normale Supérieure, Paris, France \\ 
$^3$Department of Physics, University of Illinois at Urbana-Champaign, Urbana, IL, USA \\ 
$^{\dagger}$These authors contributed equally to this publication.}


\date{\today}

\begin{abstract}
\change{One of the primary challenges in realizing large-scale quantum processors is the realization of qubit couplings that balance interaction strength, connectivity, and mode confinement. Moreover, it is very desirable for the device elements to be detachable, allowing components to be built, tested, and replaced independently. In this work, we present a microwave quantum state router, centered on parametrically driven, Josephson-junction based three-wave mixing, that realizes all-to-all couplings among four detachable quantum modules. We demonstrate coherent exchange among all four communication modes, with an average full-$\sg$ time of $\unit[764]{ns}$ and average inferred inter-module exchange fidelity of 0.969, limited by mode coherence. We also demonstrate photon transfer and pairwise entanglement between module qubits, and parallel operation of simultaneous $\sg$ exchange across the router. Our router-module architecture serves as a prototype of modular quantum computer that has great potential for enabling flexible, demountable, large-scale quantum networks of superconducting qubits and cavities.}

\end{abstract}

\maketitle

\section{Introduction}

Building quantum information processors of increasing size and complexity requires meticulous management of both the qubits' environment and qubit-qubit interactions. 
Suppressing interaction with the external environment has always been recognized as the central difficultly in maintaining coherence in a system.  In the long term, this challenge will be met by fault tolerantly encoding much smaller logical machines inside a qubit fabric. In the present so-called Noisy Intermediate-Scale Quantum (NISQ) era, we must simply run short circuits.  However, in both the short and long term, we face choices about which architecture of machine we build.  In large-scale processors based on a monolithic fabric of nearest-neighbor interactions, the circuit topology/connectivity must be carefully designed to avoid spectator qubit errors and long-range cross-talk \cite{majumder2020real, krinner2020benchmarking, dai2021calibration}. In addition, we face the challenge of fabricating all components to perform flawlessly on a single die \cite{arute2019quantum}.

\begin{figure*}
\includegraphics{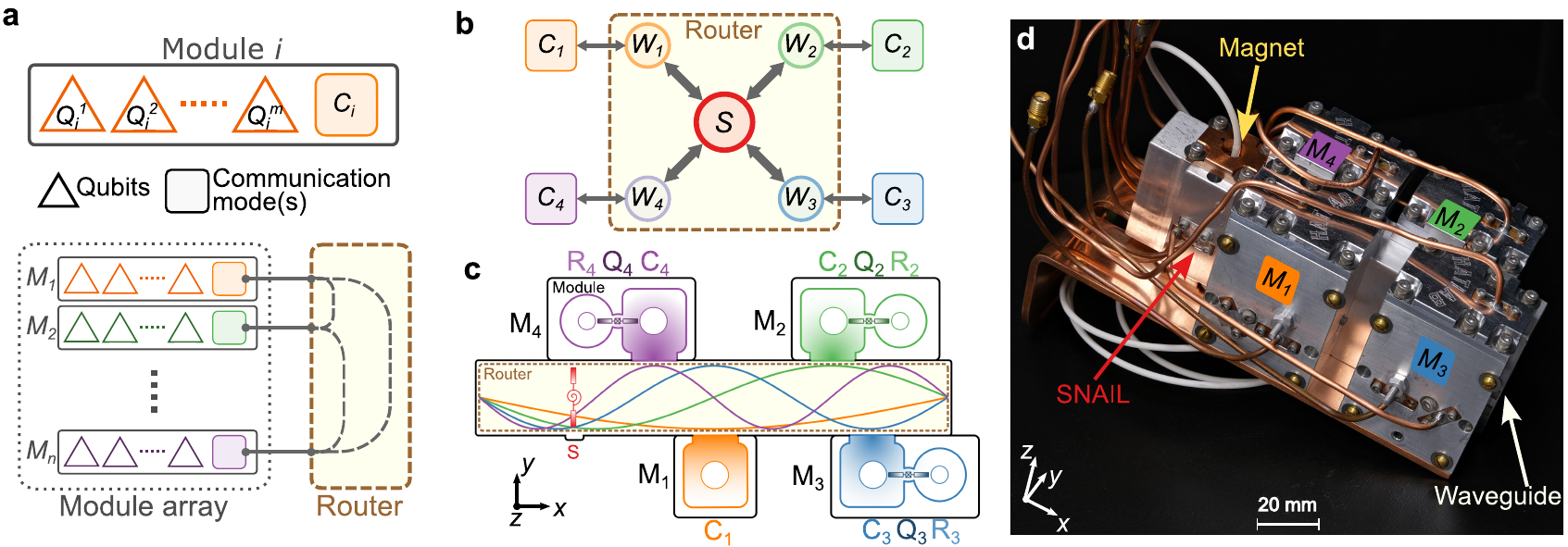}
\caption{\textbf{Schematic representation of a modular quantum computer and our prototype realization using 3D superconducting circuits.} (a) Proposed structure for a modular superconducting quantum computer, in which a number of quantum modules are connected via their communication modes to a quantum state router. (b) Coupling scheme between the router and four communication modes in our prototype device. The brown dashed square represents the router with four waveguide modes ($\textrm{W}_1-\textrm{W}_4$) and a SNAIL ($\textrm{S}$). Each waveguide mode is dispersively coupled to a single communication cavity mode ($\textrm{C}_1-\textrm{C}_4$). (c) Schematic drawing of our full system consisting of four modules and the central quantum state router. The colored curves inside the router represent the electric field (\textbf{E}) of the first four waveguide $\text{TE}_{10n} (n=1,2,3,4)$ eigenmodes.  Since each waveguide mode has a different \textbf{E}-field distribution and the SNAIL mode is detuned differently from each waveguide mode, we can place the SNAIL chip (represented in red) at a location where it has different coupling strengths $g_{\textrm{w}_i s}$, but similar hybridization strengths  $(\frac{g}{\Delta})_{\textrm{w}_is}$, to each waveguide mode. Each module (for $\textrm{M}_2$ to $\textrm{M}_4$) consists of a qubit ($\textrm{Q}_2 - \textrm{Q}_4$), a communication cavity (square shaped cavities, $\textrm{C}_2-\textrm{C}_4$) and a readout cavity (round shaped cavities, $\textrm{R}_2-\textrm{R}_4$) All the cavities used here are coaxial $\lambda/4$ cavities \cite{reagor2016quantum}. In module $\textrm{M}_1$ the qubit has been omitted. Each communication cavity is designed to have a frequency that is close to one waveguide mode and far away from the others to ensure a desired hybridization strength to the router. (d) Photograph of the assembled device.}
\label{fig:schematic}
\end{figure*}

Modular quantum systems offer a very promising alternate route to large-scale quantum computers, allowing us to sidestep many of these difficulties and instead operate using smaller, simpler quantum modules linked via quantum communication channels\cite{cirac_quantum_1997,  kimble_quantum_2008,  monroe2013scaling, monroe2016quantum}.  Such machines allow us to replace faulty components and test sub-units separately, which can greatly ease requirements for flawless fabrication while also allowing distant qubits to communicate with many fewer intermediate steps, potentially enhancing fidelity in near-term quantum processors\cite{laracuente2022short}. Moreover, the quasi-particles that cause additional bit flip and phase flip errors would also be constrained inside the module instead of propagating across the whole monolithic processor, reducing correlated errors in modular structures\cite{mcewen2022resolving, wilen2021correlated}.

The key element that determines the performance of a modular machine is its quantum communication bus. For atomic scale qubits (which form the  basis for many of the early proposals for modular quantum computing) communicating using optical-frequency states, it is infeasible to couple photons into a communication channel with very high efficiency. This loss of information precludes light from simply being transferred from module to module. Instead, one must herald instances in which transmission is successful \cite{chou2005measurement, moehring2007entanglement, ritter2012elementary, hofmann2012heralded, bernien2013heralded}. However, once light has been coupled into an optical fiber, it can be readily distributed over kilometer and longer distances, which readily supports long-range entangled state generation and distributed quantum computation\cite{wengerowsky2019entanglement, yu2020entanglement}.  In superconducting circuits, there have also been several recent demonstrations of similar measurement-based protocols\cite{roch2014observation, narla2016robust, dickel2018chip, kurpiers2019quantum}.

However, superconducting circuits can also transfer states directly. For this form of direct state exchange, we require strong, switchable couplings from module to the communication channel to enable rapid operations, low losses in the channel, and a dense, reconfigurable network of couplings among many modules\cite{monroe_large-scale_2014, linke_experimental_2017}. Realizations to date have focused on pairs of quantum modules \cite{axline2018demand, campagne2018deterministic, kurpiers2018deterministic, leung2019deterministic, magnard2020microwave} or modes in a monolithic device \cite{sirois2015coherent, mckay2016universal, noguchi2020fast, hazra2021ring, zhong_deterministic_2021}. They have utilized transmission-line based `quantum bus' communication channels and controllable module-bus couplings based on the nonlinearity of Josephson junctions or driven exchange via a driven, nonlinear coupling mode.

In this article, we propose and experimentally implement a scheme for creating a modular superconducting network, which instead creates a nonlinear `quantum state router' with fixed, dispersive couplings to individual quantum modules.  The strong, parametrically driven nonlinearity of the quantum state router allows us to only virtually occupy its modes, and thus achieve efficient operations over the router with only modest requirements for router quality. The router does not use measurement to herald entanglement between modes. Instead, operations over the bus can be thought of as direct, parametrically actuated gates between quantum modules.  The state router naturally supports all-to-all coupling among several quantum modules, and is naturally extensible to a larger modular network. We have realized a quantum state router centered on  Superconducting Nonlinear Asymmetric Inductive eLement (SNAIL)-based nonlinearity\cite{frattini20173}, and used it to operate a four module quantum processor.

\section{Results}

\textbf{Theory of router operation.}  Our proposed structure for a modular superconducting quantum computer consists of two major parts: a quantum state router and multiple modules, as shown in Figure~\ref{fig:schematic}a. Each module consists of a variable number of qubits (one in our present experiment) which have controllable, local coupling with each other. In each module, there should also be at least one ``communication'' mode which couples to both the qubits in the module and the quantum state router. This communication mode can either be a qubit or, as in this work, a long-lived harmonic oscillator which can store information for exchange over the router.

In this work, we have experimentally realized a prototype version of such a modular architecture device with 3D superconducting circuits, and have adopted several design rules to guide our efforts. First, the communication modes we use are superconducting 3D cavities rather than qubits (Figure~\ref{fig:schematic}b, c and \change{Supplementary  Figure 1}), as they accommodate multiple qubit encoding schemes including the Fock encoding used in this work, cat states\cite{vlastakis_deterministically_2013}, binomial encodings\cite{michael_new_2016}, GKP-encodings\cite{gottesman_encoding_2001}, etc.  This allows our router to be compatible with a wide array of future module designs. Second, we emphasize the ``modularity'' of our system in the additional sense that each module and the router itself exist as independent units which can operate individually, instead of the whole system forming a monolithic block. This offers a tremendous advantage in the laboratory, as defective components can be easily replaced, and the different components can be tested separately and then assembled.   Third, the router operates via  coherent photon exchange  based on parametric driving of a 3-wave-mixing Hamiltonian, in which the third-order nonlinearity is introduced by a SNAIL device.  Finally, we have designed the router to minimize both the need for precise frequency matching between router and module modes and the requirements for high-Q router elements.  To accomplish this, we couple all modes in the system dispersively.

The only nonlinear element in the router is a central SNAIL-mode $\textrm{S}$ (with corresponding annihilation operator $\hat{s}$), which is very strongly coupled to an input line for strong parametric driving, and flux biased via a nearby copper-sheathed electromagnet. As such, it has a low quality factor Q ($\sim 10,000$). The remainder of the router is composed of a rectangular, superconducting 3D waveguide, as shown in Fig.~\ref{fig:schematic}c. The first four transverse electric modes ($\text{TE}_{10i}, i=1,2,3,4$) of the waveguide $\textrm{W}_i$ (with annihilation operators $\hat{\textrm{w}}_i$) are each used as an intermediate mode coupling to both the SNAIL and a corresponding communication mode $\textrm{C}_i$ (with operator $\hat{c}_i$ in the $i$\textsuperscript{th} module). The SNAIL is flux biased to a point where its even-order nonlinear terms are negligible\cite{frattini20173, sivak_kerr-free_2019} and the third-order term is strong, resulting in the Hamiltonian of the router, which is separated into mode energies, interactions, and nonlinear terms, respectively:

\begin{align}
\label{eq:routerHamiltonian}
&\hat{\mathcal{H}}_R / \hbar  = \hat{\mathcal{H}}_{R,0} / \hbar + \hat{\mathcal{H}}_{R,\text{int}}/ \hbar + \hat{\mathcal{H}}_{R,\text{nl}}/ \hbar\\
    & \quad =\left[\omega_s \linmode{s}{} + \sum_{i} \omega_{\textrm{w}_i} \linmode{\textrm{w}}{i}\right] +\left[ \sum_{i} g_{\textrm{w}_{i}s} (\hat{\textrm{w}}_i^\dagger \hat{s} + \hat{\textrm{w}}_i \hat{s}^\dagger)\right] \nonumber \\ 
    &\quad + \left[g_{sss} (\hat{s} + \hat{s}^\dagger)^3\right]. \nonumber
\end{align}

The waveguide modes are naturally orthogonal; each is coupled to the SNAIL with strength $g_{\textrm{w}_i s}$, and $g_{sss}$ is the strength of the SNAIL's third-order term. We parametrically drive photon exchange between a pair of waveguide modes by driving the SNAIL at the difference of their frequencies.  This scheme has been long used in parametric amplifiers and circulators, where it goes by the name `noiseless photon conversion' \cite{bergeal_analog_2010, sliwa_reconfigurable_2015, lecocq2017nonreciprocal}. To have independently controllable couplings, we have chosen the SNAIL frequency and waveguide dimensions so that all mode frequencies and frequency differences are unique, with all difference frequencies below the lowest mode frequency (see \change{Supplementary Table 1, Supplementary Figure 4}).   

As all frequencies are widely separated, we can rediagonalize the system to eliminate the interaction term, slightly shifting all mode frequencies and definitions (for simplicity's sake, we omit any change of variable representation for the new, hybrid eigenmodes), and inducing all possible self- and cross-three-wave couplings among the waveguide modes and SNAIL. This is analogous to common techniques used in circuit QED\cite{nigg_black-box_2012, minev_energy-participation_2021}, with a third- rather than fourth-order nonlinearity. Retaining only the parametric coupling terms we will use in the router, which is safe as long as all other processes are well separated from any desired process in frequency, we write the effective Hamiltonian of the router as

\begin{equation}
\label{eq:router3waveMixing}
\mathcal{\hat{H}}^{\text{eff}}_{R}/ \hbar=\mathcal{\hat{H}}_{R,0}/\hbar+ \sum_{i\neq j} g^\text{eff}_{\textrm{w}_i \textrm{w}_j s} (\what_i^\dagger \what_j \hat{s} + \what_i \what_j^\dagger \hat{s}^\dagger).    
\end{equation}

The effective three-wave interaction strengths are given by $g^\text{eff}_{\textrm{w}_i \textrm{w}_j s} \approx 6 g_{sss} (\frac{g}{\Delta})_{\textrm{w}_is} (\frac{g}{\Delta})_{\textrm{w}_js}$, where $\Delta_{\textrm{w}_i s} = \omega_{\textrm{w}_i} - \omega_s$. We note that in our experiment, we never directly populate these waveguide modes or drive their difference frequencies. Instead, these terms serve as a `scaffold' in the router to create similar terms among the module communication modes, as detailed below. The hybridization strengths $(\frac{g}{\Delta})_{\textrm{w}_is}$ are key parameters, as they both limit the eventual parametric coupling strengths and determine how much longer-lived the waveguide modes can be compared to the low-Q SNAIL mode.

Next, we combine our router with the modules' communication modes.  As shown in Fig.~\ref{fig:schematic}b, we accomplish this by creating four modules, each containing one mode with a frequency near one of the router's waveguide modes, and coupling to the router via an aperture in their shared wall. This coupling is deliberately dispersive, with the strength controlled by a combination of waveguide-communication mode detuning, coupling aperture size, and placement along the router's length. Assuming each cavity is only coupled to its “adjacent” waveguide mode, the router plus communication mode Hamiltonian is written as:
\begin{align}
\label{eq:rcHamiltonian}
    &\hat{\mathcal{H}}_{RC}/\hbar = \Hamiltonian_R /\hbar + \Hamiltonian_{C,0}/\hbar + \Hamiltonian_{RC,\text{int}} /\hbar \\
    &\quad=\Hamiltonian_R /\hbar  + \left[ \sum_{i} \omega_{c_i} \linmode{c}{i} \right] + 
    \left[ \sum_{i}  g_{c_i \textrm{w}_i} (\hat{c}_i^\dagger \what_i + \hat{c}_i \what_i^\dagger)\right]. \nonumber
\end{align}
The second and third terms denote the communication mode's  energy and the communication mode-waveguide mode interactions, respectively. As before, we diagonalize this Hamiltonian to eliminate the direct interactions among the modes without changing variable representation, and neglect all but the cavity-cavity third-order interactions to find the new effective Hamiltonian for the composite router plus communication modes system:

\begin{equation}
\label{eq:cavity3waveMixing}
\mathcal{\hat{H}}^{\text{eff}}_{RC}/ \hbar=\mathcal{\hat{H}}_{R,0}/\hbar+ \mathcal{\hat{H}}_{C,0}/\hbar+\sum_{i\neq j} g^\text{eff}_{c_i c_j s} (\chat_i^\dagger \chat_j \hat{s} + \chat_i \chat_j^\dagger \hat{s}^\dagger).  
\end{equation}
The resulting effective three-body interaction strength is $g_{c_i c_j s}^\text{eff} \approx g_{\textrm{w}_i \textrm{w}_j s}^\text{eff} (\frac{g}{\Delta})_{c_i \textrm{w}_i} (\frac{g}{\Delta})_{c_j \textrm{w}_j}$. 
 
 The use of a network of hybridization, linking the cavity modes to the central SNAIL via intermediate cavity modes comes with advantages: Since the SNAIL pump port is physically separated from the communication modes in two different metal bodies, we can assume no direct coupling between them.  Thus, by choosing each dispersive coupling $g/\Delta \simeq 0.1$, the weakly hybridized communication modes can live up to $10^4$ times longer than the SNAIL mode and $100$ times longer than the waveguide modes, greatly decreasing the need for long lifetime components in the router.

Moreover, dispersive couplings and parametric driving are insensitive to modest errors in mode frequencies, further reducing the need for precision fabrication, unlike photon exchange techniques based on resonant mode couplings\cite{bialczak_fast_2011, leung2019deterministic}. While it is certainly possible to remove the intermediate waveguide modes in a monolithic version of our design, this comes with both greatly reduced flexibility in combining disparate elements and more stringent requirements for the SNAIL's lifetime.  

In operation, the parametrically driven two-body exchange rate, for example between modes $\textrm{C}_i$ and $\textrm{C}_j$, is $\sqrt{n_s} g_{c_i c_j s}^{\mathrm{eff}}$, where $n_s$ is the pump strength expressed as a photon number (see \change{Supplementary Method 1}).  It is here that we find the price for our hybridization network: the effective three body coupling has been greatly reduced ($g_{c_i c_j s}^{\mathrm{eff}} \simeq 6\times10^{-4} g_{sss}$). To achieve rapid gates with feasible pump strengths, we must both engineer $g_{sss}$ to be large and carefully design the pump line to tolerate very strong drives to compensate for this dilution of nonlinearity. The former is controlled by the SNAIL \change{(circuit diagram shown in Supplementary Figure 2c)} parameters including Josephson inductance of the small junction ($L_J$), capacitance ($C$) and junction inductance ratio ($\alpha$). Unlike in amplifier designs where arrays of SNAIL loops were made to suppress Kerr non-linearity\cite{sivak_kerr-free_2019, frattini2018optimizing}, here we choose to make a single-loop SNAIL  with relatively large $\alpha$ to give stronger $g_{sss}$\cite{frattini20173}. Specifically, for the device we used in this experiment, we design $L_J=3.44~\text{nH}, C=0.456~\text{pF and}~\alpha = 0.28$.

The pump line design aims to deliver sufficient power to the SNAIL mode without overheating the fridge. This requires strong coupling between the pump port and the SNAIL mode, but this can also Purcell limit the lifetimes of the waveguide modes, thereby limiting the communication cavity modes' lifetimes.  However, the off-resonance nature of our parametric pumping scheme actually allows us to separate these two constraints in the frequency domain. Specifically, all the possible pumping frequencies we used here are below all the mode frequencies (\change{Supplementary Figure 4}). This allows us to use a reflective low-pass filter (LPF) on the SNAIL pump port (as shown in Supplementary Figure 3) that protects the modes in our device while allowing high-power, low-frequency pumps to pass. Moreover, unlike on-resonance driving schemes that require large amounts of attenuation (usually 20-30 dB) on the mixing chamber (MC) plate to reduce stray photons in the drive lines, the reflective LPF used here also generates much less heat, which also gives more tolerance to strong external pumps, thereby making the parametric pumping scheme even more promising in realizing large-scale/multiplexed qubit controls.

It is also important to note that Eq.~\ref{eq:cavity3waveMixing} represents only our desired coupling terms. In practice, all modes inherit both self- and cross-three-wave mixing terms from the SNAIL.  Of particular concern are couplings between a cavity and non-adjacent waveguide modes (e.g.\ $\textrm{C}_3$ to $\textrm{W}_4$). To avoid potential issues, we choose a minimum waveguide to cavity spacing of $\unit[\sim100]{MHz}$ to suppress cross-talk with these couplings.  This difference frequency is comparable to the anharmonicity of transmon qubits; similarly, we can use variants of the DRAG\cite{motzoi_simple_2009} technique to drive rapid $\sg$ gates among the communication modes without leakage to unwanted modes (see \change{Supplementary Discussion 1}).

\begin{figure}[ht]
\includegraphics[]{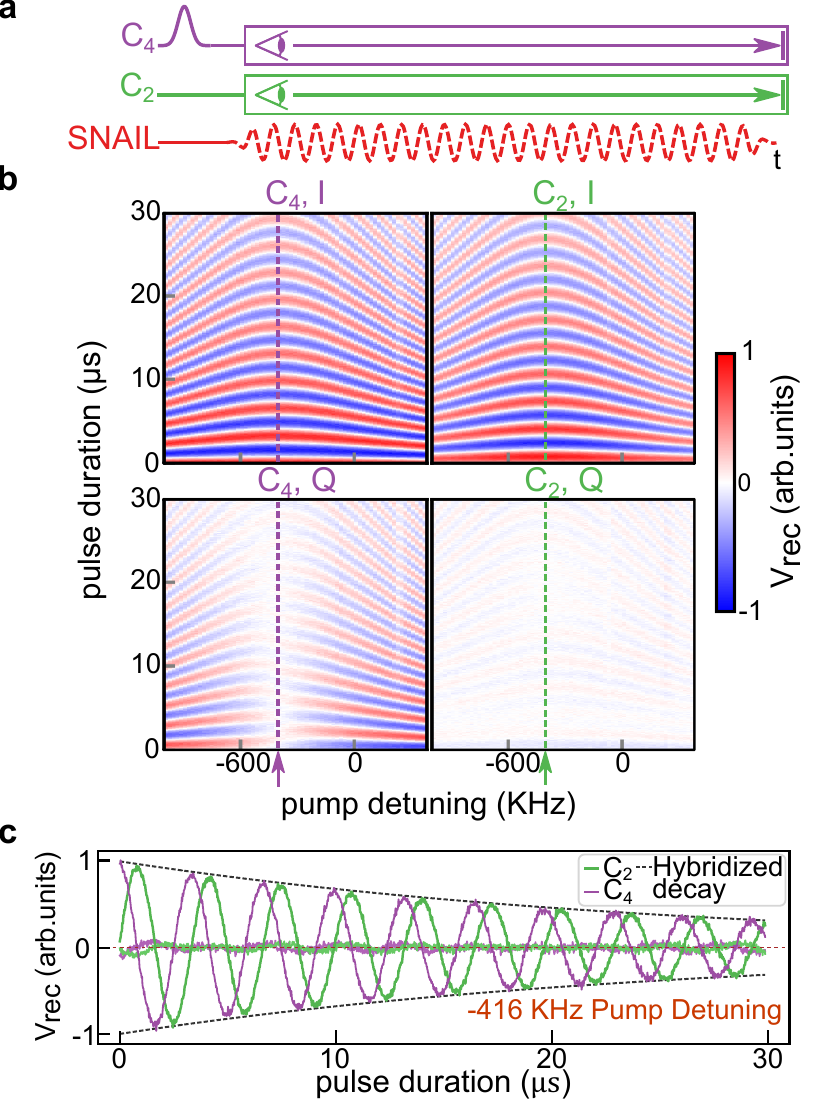}
\caption{\textbf{Coherent state exchange between communication cavities.} (a) Pulse sequence of the swap experiment.  We begin by displacing one cavity to create a coherent state, which we then swap between a pair of cavities by applying a parametric drive to the SNAIL.  We continuously monitor the I-Q voltage in each cavity during the swap process.
(b) In-phase and quadrature received voltage from the two cavities versus pulse duration and pump detuning from the nominal difference frequency. The dashed vertical line denotes the optimal detuning frequency for full photon exchange.
(c) Line-cut of (b) at the optimal full-swap detuning. The grey dashed envelope represents the hybridized $T_2$ decoherence  of the coupled systems, given by $\exp(- \bar{\Gamma}_2 t)$, where $\bar{\Gamma}_2 = (1/T_{2, \textrm{C}_2} + 1/T_{2, \textrm{C}_4})/2$ is the averaged decoherence rate of the two cavities involved (here $\textrm{C}_2$ and $\textrm{C}_4$), as the photon being exchanged spends half of its time in each cavity.}
\label{fig:swap}
\end{figure}

\textbf{Basic router characterization with coherent states.} For initial experiments, we connect our router to four simple modules, but omit the module qubits.  Each communication mode is driven and characterized via an under-coupled probe port whose induced relaxation rate is much smaller than the mode's internal loss rate.  Figure~\ref{fig:swap}a shows an experimental pulse sequence for swapping coherent states between the module communication modes $\textrm{C}_2$ and $\textrm{C}_4$. First, a short on-resonance drive is applied to  $\textrm{C}_4$ through the weakly coupled port, which creates a coherent state in this cavity. Then, a pump tone is applied to the SNAIL mode near the $\textrm{C}_2-\textrm{C}_4$ difference frequency $\omega_p = \omega_{c_4} - \omega_{c_2} + \delta$, where $\delta$ is the pump detuning relative to the measured frequency difference between the two cavity modes.  Meanwhile, the light in these two cavities is monitored by receiving the I-Q signal leaking out from each cavity's probe port. The amplitude of the coherence state inside the cavity can then be inferred by demodulating the signal at the corresponding cavity frequencies. By sweeping the applied pump frequency and time, we can determine both the swap rate and resonant condition for pumping.  

The experiment results are shown in Fig.~\ref{fig:swap}b, c.  There is a good agreement between the envelope of the swap trace (green and purple lines) and the hybridized decay trace, indicating that the state is only swapping between these two cavities without leaking into other modes,  and that the  fidelity of state exchange is mainly limited by the lifetime of these two cavities. This same experiment is then performed for the six possible pairs of the four communication modes. We find the fastest full-swap time to be $\unit[375]{ns}$, the slowest $\unit[1248]{ns}$, and an average swap time of $\unit[\sim764]{ns}$ (see \change{Supplementary Table 2}). For each  pair, the maximum gate speed is measured by increasing the pumping power until we see the mode coherence times substantially differ from their undriven values.  On average, the pump frequency required to fully swap light between the two cavities is detuned by several hundred kHz ($\unit[-416]{kHz}$ for the data in Fig.~\ref{fig:swap}). We attribute this to a combination of SNAIL- and communication-mode static and dynamic Kerr effects; that is, the communication-mode frequencies are shifted due to the off-resonance pump on the SNAIL mode. In our pulses, we are only sensitive to changes in the communication modes' frequency difference, so shifts can be positive/negative/zero depending on whether the two modes shift closer together or further apart. This has strong parallels to saturation effects in parametric amplifiers, where the amplifier modes shift with stronger pumping, which complicates amplifier bias and can lead to limitations in power handling in the amplifier as bigger input signals shift the amplifier away from its small-signal frequency \cite{frattini2018optimizing, sivak_kerr-free_2019, planat_understanding_2019,liu_optimizing_2020}.  Fortunately, in the case of parametrically controlled \sg gate the primary consequence is simply that we must track these shifts and their effects on the qubits' phases in our control electronics.

\begin{figure}[h]
\includegraphics[]{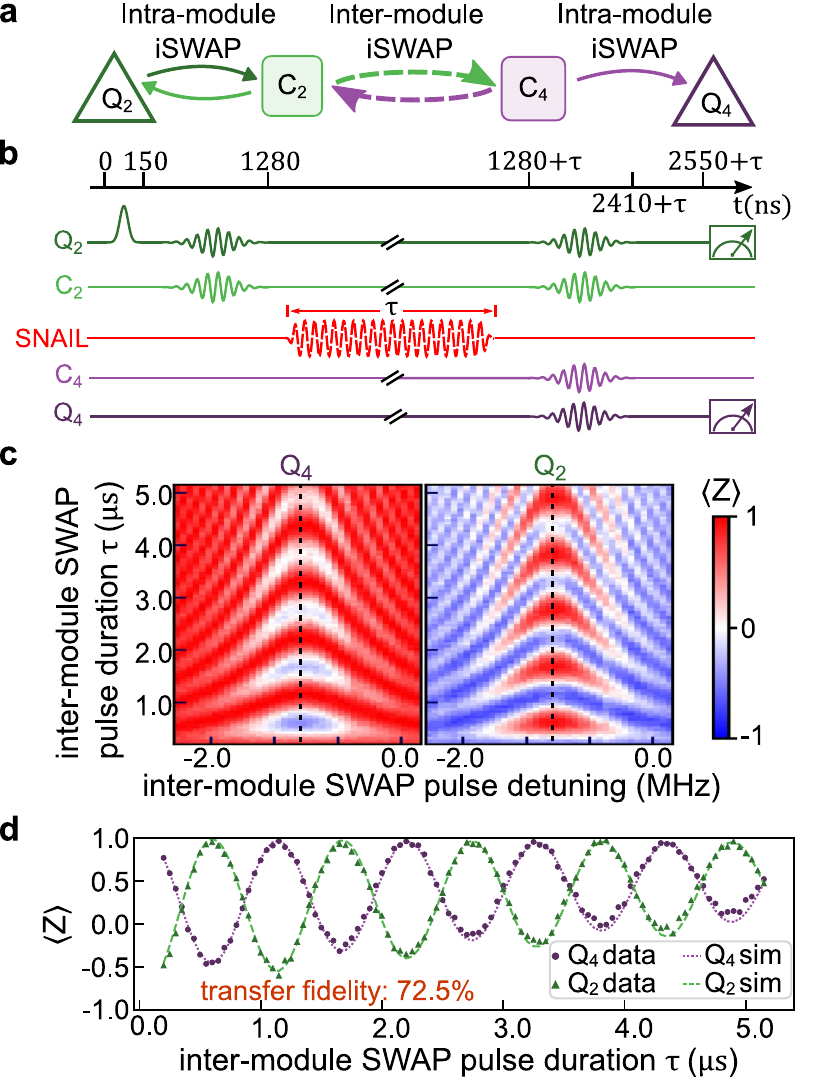}
\caption{\textbf{Fock state swap experiment between remote qubits.} (a) Illustration of the photon swap protocol, in which a photon originating in $\textrm{Q}_2$ is fully swapped to $\textrm{C}_2$, then depending on the variable inter-module pulse duration, routed to $\textrm{Q}_4$ or returned to $\textrm{Q}_2$. (b) Experiment pulse sequence.  A photon is created in $\textrm{Q}_2$, then swapped to $\textrm{C}_2$.  Next, it is swapped (or not) to $\textrm{C}_4$ with a variable duration, inter-module $\sg$ pulse. Finally, the light in $\textrm{C}_{2,4}$ is routed via further intra-module $\sg$s to their respective qubits, which are then measured. The upper black bar indicates the total experimental duration with $\tau$ describing the variable, SNAIL actuated inter-module $\sg$. (c) Measurement result of $\textrm{Q}_2$ and $\textrm{Q}_4$ for different SNAIL pump detuning and duration. Here, the color of the 2D sweep indicates the measurement along the qubits' $z$-axis. (d) A cut of the swap data along the dotted line indicated in (c). The green triangles and purple circles are $\textrm{Q}_2$ and $\textrm{Q}_4$ data, respectively, and the dashed lines are the corresponding simulation results.}
\label{fig:QQSWAP}
\end{figure}

\textbf{Full device operation with single-qubit modules} Next, we add the transmon qubits to complete modules 2-4 (module 1's qubit is omitted), and perform full intra-module and inter-module operations across the device. Each single-qubit module consists of one communication cavity $\textrm{C}_i$, one transmon $\textrm{Q}_i$, and one readout cavity $\textrm{R}_i$. The device layout is shown schematically in Fig.~\ref{fig:schematic}c.  For simplicity, our qubit states throughout the system are the Fock states $\ket{0}$ and $\ket{1}$, although the communication modes could in principle support a variety of more complex encodings. Measured coherence rates ($T_1$, $T_{2R}$, and $T_{2E}$) can be found in \change{Supplementary Table 1}. We perform intra-module gates between the qubit and cavity in each module using a doubly-driven parametric photon exchange process (see \onlinecite{narla2016robust},\onlinecite{axline2018demand},  \onlinecite{burkhart2020error} and  \change{Supplementary Method 2} for details), which are indicated as paired drives in Fig.~\ref{fig:QQSWAP}b. Here, the communication cavities serve as intermediary modes that only store  (but not compute on) quantum states, and enable the photon exchange between modules via the router controlled \sg~gates.

In the simple algorithms that follow, we refer to the operation of these exchange interactions as variations of the $\sg$ gate, as is typical for gates based on coherent photon exchange\cite{sirois2015coherent, narla2016robust, mckay2016universal, burkhart2020error, zhong_deterministic_2021, sung2021realization}. We exclusively use these gates to swap coherent states or Fock states fully from a source cavity to a formerly empty target cavity. In this scenario, the gates act as a combination of SWAP and $z$-rotation for both Fock and coherent states. However, this analogy breaks down for both intra- and inter-module exchange of Fock states between a qubit and cavity and a pair of cavities, when we consider arbitrary pulse lengths and certain joint qubit-cavity or cavity-cavity Fock states (e.g.\ $\ket{1,1}$).  For this reason, some researchers choose to refer to such gates between pairs of cavities as `beam splitters'\cite{burkhart2020error, gao2018programmable, pfaff2017controlled} for their obvious resemblance to the optical component of the same name.  This analogy, however, fails for our qubit-cavity interactions, so we choose instead to refer to these gates via the exponent which determines their unitary relative to a full $\sg$ gate, i.e., a $\rtsg$ is described by $U_{\sg}^\frac{1}{2}$. Because our protocols only swap light into empty qubits/cavities, states containing two or more photons are never occupied. As such, this inexact analogy yields both a simple graphical and conceptual picture of our gates, as well as correct intuition about the system's evolution during our protocols. This issue, however, must obviously be revisited for alternate qubit encoding choices, and when both send-modes and receive-modes are in arbitrary states.  For further discussion, see \change{Supplementary Discussion 2.}

We next use the module transmons and intra-module $\sg$ operations to swap Fock states across the router, transferring single photons between distant qubits as shown in  Fig.~\ref{fig:QQSWAP}a, b. The protocol begins with all qubits and cavities prepared in their ground states. A $R_x(\mathrm{\pi})$ pulse is first applied to $\textrm{Q}_2$ which brings it to the excited state. Second, an intra-module $\sg$ gate is performed between $\textrm{Q}_2$ and $\textrm{C}_2$. This fully swaps the excitation from $\textrm{Q}_2$ to $\textrm{C}_2$. Third, the photon is swapped between $\textrm{C}_2$ and $\textrm{C}_4$ across the router by pumping on the SNAIL mode, just as demonstrated in Fig.~\ref{fig:swap}. The SNAIL pump duration is varied, which results in an effective Rabi oscillation between the two qubits when the protocol is completed.  Finally, we apply two more intra-module $\sg$ gates, $\textrm{C}_2$ to $\textrm{Q}_2$ and $\textrm{C}_4$ to $\textrm{Q}_4$. This fully transfers the states of $\textrm{C}_2$ and $\textrm{C}_4$ to their respective module qubits, which are then measured simultaneously using dispersive readout of the readout ($\textrm{R}$) modes. The results are shown in Fig.~\ref{fig:QQSWAP}c and ~\ref{fig:QQSWAP}d. The transfer fidelity between $\textrm{Q}_4$ and $\textrm{Q}_2$ is $72.5 \pm 1.17~\%$.  We perform Lindblad master equation simulations assuming ideal interactions, with the only defect being all modes' measured coherences  \change{(see the Methods section)}; the simulation results (dotted curves) show a good quantitative agreement with our data, indicating that, as with coherent state operation, the primary fidelity limit in our system is the ratio of gate time to our modes' coherence times. The uncertainty given for the Fock state transfer fidelity, and all following quoted fidelities, is calculated following the `bootstrap method' in Ref. \onlinecite{newman1999monte, young_everything_2014}, \change{and is explained in the Methods section}.  \change{In the above calculated fidelities, no correction is applied for State Preparation and Measurement (SPAM) errors.}  Details of experimentally determined SPAM errors can be found in \change{Supplementary Table 1.}

\begin{figure}[h]
\includegraphics[]{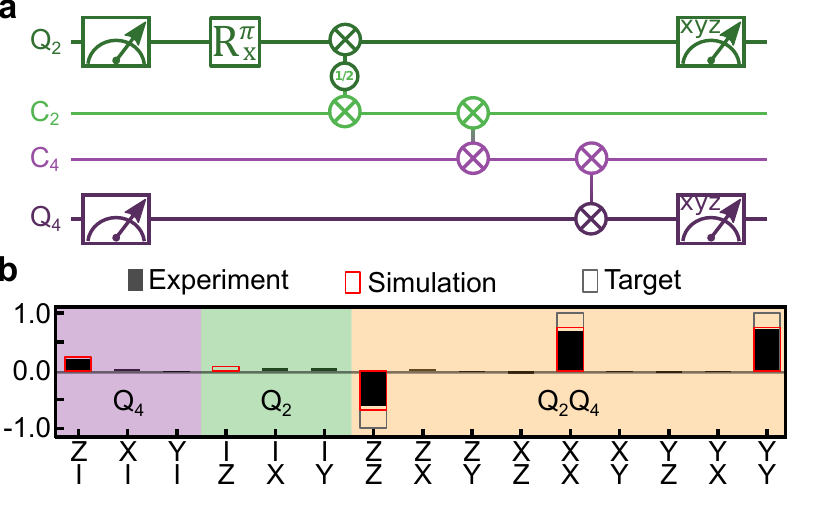}
\caption{\textbf{Inter-module Bell state generation.} (a) Quantum circuit for generating a Bell state between $\textrm{Q}_2$ and $\textrm{Q}_4$. Entanglement is first generated between $\textrm{Q}_2$ and $\textrm{C}_2$ using a $\rtsg$ gate, then the cavity component is moved to $\textrm{Q}_4$ using two full-$\sg$ gates. (b) Tomography of the joint $\textrm{Q}_2, \textrm{Q}_4$ Bloch vector, in which each bar represents a joint measurement of the two qubits in the basis indicated ($I$ indicates no measurement). Here, the black bars indicate the experimental result, the red rectangles are master-equation simulation results, and the gray rectangles represent the pure Bell state. The fidelity to the target Bell state $\frac{1}{\sqrt{2}}\left(\left| 01 \right \rangle + \left| 10 \right \rangle\right)$ is $ 76.9 \pm 0.76\% $, which agrees very well with the simulation prediction of $77.2\%$.}
\label{fig:bellstate}
\end{figure}

\textbf{Inter-module Bell state generation.}  
Next, we utilize a $\rtsg$ gate, created by shortening the first intra-module $\sg$ gate from Fig.~\ref{fig:QQSWAP} by close to 1/2 in duration, to create inter-module Bell states. The $\rtsg$ has the effect of taking the single photon in the qubit and coherently `sharing' it between the qubit and cavity, creating a Bell state between the two modes.  Overall, the sequence first creates a Bell-pair inside a module, then shifts the communication cavity's component to a qubit in a second module. The quantum circuit is shown in Fig.~\ref{fig:bellstate}a. Tomography is performed on both qubits, while the communication cavities are not measured. The measurement results are shown in Fig.~\ref{fig:bellstate}b. From this tomographic data, we can reconstruct the density matrix of $\textrm{Q}_2$ and $\textrm{Q}_4$, and find we achieve a Bell fidelity of $76.9 \pm 0.76~\%$. The same experiment is performed on the other two qubit pairs $\textrm{Q}_2-\textrm{Q}_3$ and $\textrm{Q}_4-\textrm{Q}_3$) with fidelities of $58.7 \pm 2.40~\%$ and $68.2 \pm 0.83~\%$, respectively. The results were again compared with Lindblad master equation simulations (red rectangles in Fig.~\ref{fig:bellstate}b), and show that the dominant source of infidelity remains the modes' lifetimes. In addition, we attempted a GHZ state preparation experiment between all three qubits in the modules using a similar scheme. The result is discussed in \change{Supplementary Discussion 4}.

\begin{figure}[h]
\includegraphics[]{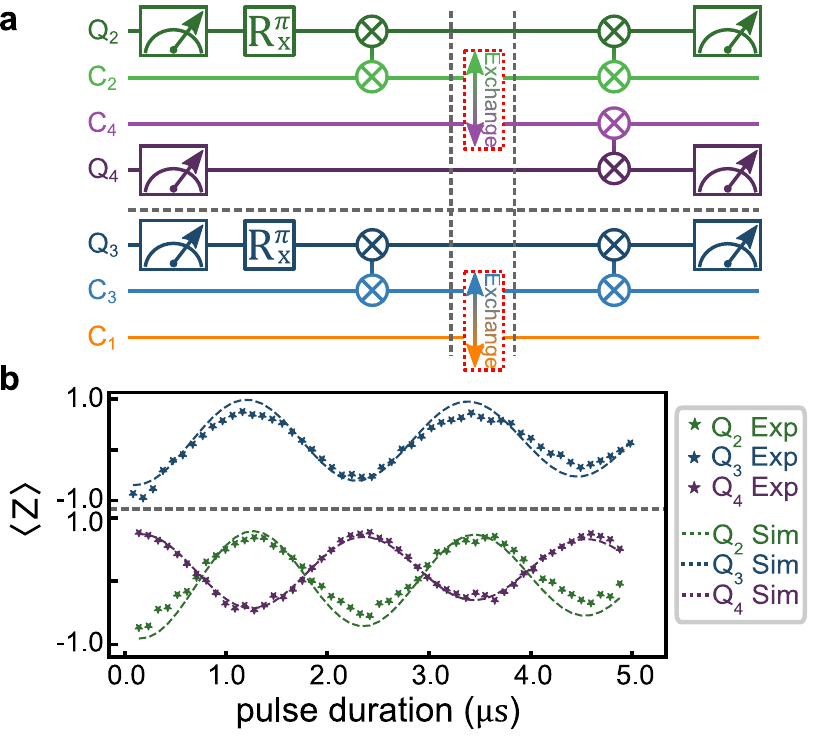}
\caption{\textbf{Parallel photon exchange experiment.} (a) Gate sequence for parallel photon exchange over the router. (b) Photon population of all three qubits vs. router swap time. The dots are experimental results, and the corresponding dashed lines are simulation results.}
\label{fig:parallel}
\end{figure}

\textbf{Parallel operations.} Another advantage of our architecture is that we can drive multiple parametric operations in the router simultaneously, which enables parallel operation and efficient ways to create entanglement. We demonstrate the simplest implementation of parallel operations by swapping light between two pairs of modules simultaneously. Here, $\textrm{M}_2$ and $\textrm{M}_4$ are treated as one sub-system, while $\textrm{M}_3$ and $\textrm{M}_1$ form the second one. We swap a photon from $\textrm{Q}_2$ to $\textrm{Q}_4$ and $\textrm{Q}_3$ to $\textrm{C}_1$  across the router simultaneously. The gate sequence is shown in Fig.~\ref{fig:parallel}a.  The two cross-module swap interactions, $\textrm{C}_2 - \textrm{C}_4$ and $\textrm{C}_3 - \textrm{C}_1$, are turned on simultaneously by pumping the SNAIL mode at the two difference frequencies using a room-temperature combiner. The SNAIL pumps are applied for a variable period. The protocol concludes with SWAP gates between all cavity-qubit pairs and measurement of all qubits. 

The results (Fig.~\ref{fig:parallel}b) show that Fock states can swap between both pairs of modules simultaneously without interference or enhanced relaxation, as shown by comparison to master equation simulations. The drive frequencies for parallel swap processes in the router need frequency adjustments on the order of ${\sim}\unit[100]{kHz}$ compared to the single $\sg$ case, which we attribute to dynamic and static cross-Kerr effects due to the paired SNAIL drives. We reduce the pump strengths, slowing the gates from $\unit[600]{ns}$ to $\unit[1300]{ns}$, as we observe additional decoherence when running two parallel processes at maximum pump strength. We do not believe this is a fundamental limitation, but can be improved in future experiment by optimized SNAIL and router design.

As further proof of the quantum coherence of parallel operations in the router, we repeat the Bell state generation protocol between $\textrm{Q}_2$ and $\textrm{Q}_4$ with the $\textrm{M}_1-\textrm{M}_3$ $\sg$  activated in parallel. Again, the pump strengths are decreased, slowing the inter-module swap time. We achieve a Bell state fidelity of $68.1 \pm 0.79~\%$, while the simulated fidelity is $68.4~\%$. Here, the decrease of fidelity compared to the single Bell state generation process (which has a fidelity of $76.9 \pm 0.76~\%$) is due to the longer gate time used for the $\textrm{C}_2-\textrm{C}_4$ $\sg$ in the presence of a parallel $\sg$  operation (see details in \change{Supplementary Discussion 5}).

\begin{figure}[h]
\includegraphics[]{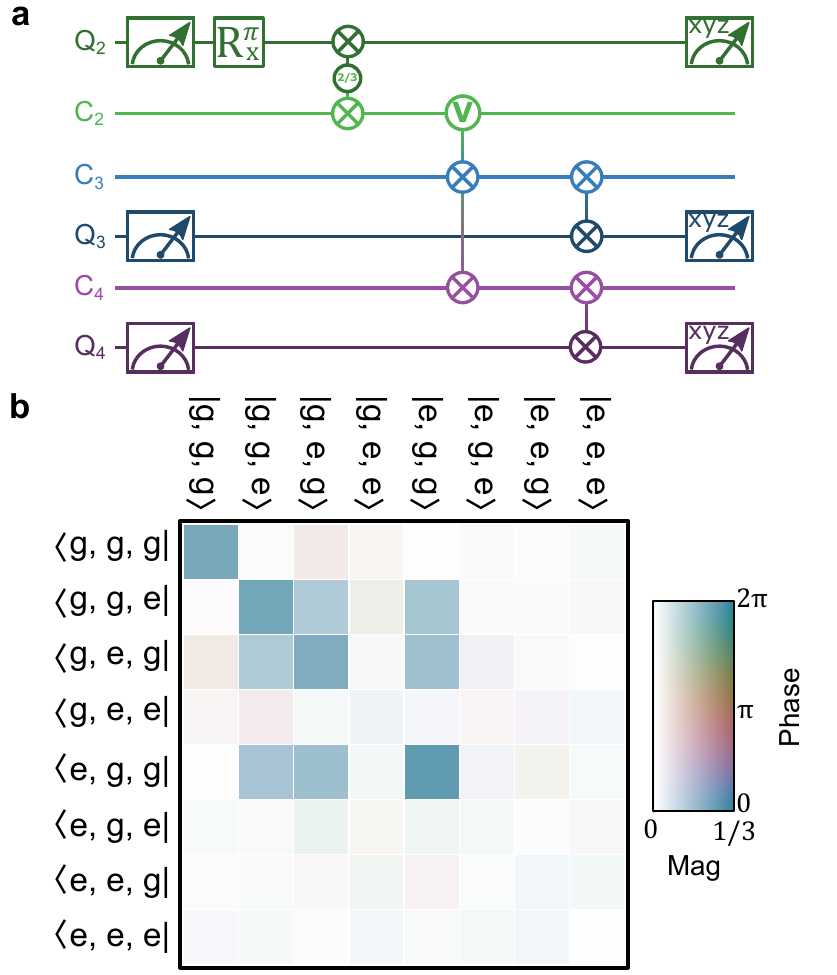}
\caption{\textbf{W state generation with `V-$\sg$' gate.} (a) W state generation  pulse sequence.  Together,  the ($\sg$)$^{2/3}$ and `V-$\sg$' gates create a W state distributed across $\textrm{Q}_2$, $\textrm{C}_3$, and $\textrm{C}_4$.  The subsequent $\sg$s redirect the latter two components to $\textrm{Q}_3$ and $\textrm{Q}_4$, respectively. (b) W state generation density matrix reconstructed from tomography. Each element in the density matrix is represented by a color using the Hue-Chroma-Luminance (HCL) color scheme. The amplitude of each element is mapped linearly to the Chroma and Luminance of the color, and the phase (from 0 to 2$\mathrm{\pi}$) is mapped linearly to the Hue value. This color mapping scheme has the property that elements of the same amplitude are perceived equally by the human eye, so that the small magnitudes fades into the white background to avoid drawing the eye to small, noisy matrix elements. The observed fidelity of the state is $53.4 \pm 2.56~\%$.}
\label{fig:vswap}
\end{figure}

\textbf{Multi-parametric gate experiment.} To demonstrate further capabilities of our system, we also explored the use of two simultaneous swap processes that link one `source' cavity to two `target' cavities. We refer to such a processes as a `V-$\sg$'. This form of swap, for a certain duration, empties the source cavity, coherently and symmetrically swapping its contents into the target cavities.  By combining the V-$\sg$ with a ($\sg$)$^{2/3}$ gate (which is realized by turning on the $\textrm{Q}_2 - \textrm{C}_2$ exchange interaction for $t = \arctan{(\sqrt{2})}/g^{\text{eff}}$) as shown in Fig.~\ref{fig:vswap}a, we can take a single photon from $\textrm{Q}_2$ and create a W-state shared among the three designated modules. We achieve a fidelity of $53.4 \pm 2.56~\%$ for this state. (see state reconstruction in Fig.~\ref{fig:vswap}b). For further discussion see \change{Supplementary Discussion 3}.

Currently, the utility of the above multi-parametric gates is limited by the slowdown of the gate times compared to individual $\sg$s. However, we believe this kind of multi-parametrically-pumped process should be further investigated, as it could be used to generate other multi-qubit gates in one step. Given the overhead in composing a multi-qubit gate from a series of two-qubit and single-qubit gates (for example, a Toffoli gate can be decomposed into 6 C-NOTs), performing these multi-parametric gates could give better performance in terms of gate fidelity by shortening the overall sequence time/gate count, even if operating at a lower rate. 

We also note that in the above multi-parametric experiments, we observe no indication of fridge heating despite two strong pumps being applied to the SNAIL. As discussed earlier in section II, the replacement of attenuation with the reflective LPF at the MC plate gives the parametric pumping scheme an advantage of 20-30 dB in fridge heating tolerance for the same circulating powers at base. 

In the above discussion, we have listed only state fidelities of combined intra- and inter-module operations. Although our current device setup does not support tomography on the communication modes, the good agreement between our experiment results and the Lindblad master equation simulations (which consider only the measured $T_1$ decay and $T_{\phi}$ dephasing of the involved modes) indicates that our inter-module photon exchange fidelity is only limited by the mode coherence times and the duration of gate operations in the pulse sequence, more importantly, our parametric pumping tone doesn't introduce extra dephasing on any of the modes in the system. Thus, we can estimate the performance of the router itself by considering the gate time ($T_{gate}^{i,j}$) and the averaged decoherence rate ($\bar{\Gamma}_2$) of each communication cavity pair \cite{gao2018programmable} $\textrm{C}_i$ and $\textrm{C}_j$, i.e. $F_{\sg}^{i,j} \simeq 1-  \bar{\Gamma}_2^{i,j}*T_{gate}^{i,j}$. Using the values listed in \change{Supplementary Table 1 and 3}, we calculate our best \sg exchange fidelity $F_{\sg}^{1,4} = 98.2\% $, the worst $F_{\sg}^{2,3} = 94.7\%$, and the average $F_{\sg}^{avg} = 96.9\% $.

\section{Discussion}
We have demonstrated a coherent quantum state router for microwave photons and used it to realize all-to-all couplings among four detachable quantum modules. The full device serves as a prototype demonstration of a modular structure superconducting quantum processor. A key feature is the use of a SNAIL mode to create three-wave couplings in the router itself, rather than relying on nonlinear couplers embedded in each module.  The router enables us to create all-to-all couplings among a set of quantum modules, to parametrically drive gates between the communication modes of those modules, and even to create three-qubit and perform parallel $\sg$ operations between multiple pairs of communication modes by applying multiple, simultaneous parametric drives.  

The current device's performance ($F_{\text{avg}}=0.969$ for gates involving the router) is limited primarily by the qubit/cavity lifetimes involved, though the limitation is primarily in the modules themselves and due to imperfect quantum engineering.  Other recent implementations of similar quantum modules\cite{burkhart2020error, axline2018demand} have achieved much higher coherence time, with qubit and cavity modes in the $\unit[100-1000]{\mu s}$ range.

With modest improvements in the lifetime of our waveguide modes to  $\sim \unit[10]{\mu s}$, our router will be able to provide sub-microsecond, very high fidelity gates between millisecond-scale communication cavities. One way this can be achieved is by retracting the SNAIL and its related lossy elements (i.e. the bias magnet and the pump port) into a coupling tube\cite{wang2016schrodinger, axline2018building}. The tube works as a waveguide with a high cut-off frequency that limits the direct coupling between our waveguide modes and the lossy elements. The SNAIL can then maintain strong coupling to the waveguide modes with its antenna sticking into the waveguide. Such a design also has the advantage of coupling a single SNAIL to multiple router elements, allowing us create inter-router operations. Through the integration of such inter-router connections and the expansion of multi-qubit modules, the structure demonstrated here can be readily expanded to build a scalable, modular network of superconducting qubits \cite{mckinney2022co}. 

Another important source of losses for the waveguide modes is the seam loss at the joint between the waveguide and the communication cavity modes. In the device reported here, these seams were sealed with indium wires, with the hope of forming a superconducting gasket between the two aluminum bodies. In our follow-on experiments, we have found that flat, polished aluminum-aluminum surface contact can give much lower loss than indium wire sealing (see discussion of improving seam losses in \cite{lei2020high}), and new devices machined using this method have shown waveguide lifetimes of  hundreds of microseconds without the SNAIL chip.

One vital question requires further research: How fast can we ultimately drive gates in this system?  A straightforward route is to further increase the waveguide mode lifetimes in the router. We can then increase the dispersive coupling strength to their respective communication modes without decreasing the communication mode lifetimes.  Doubling our current average coupling to $g/\Delta=0.2$ will immediately push the average gate time to $\unit[\sim100]{ns}$.  We must also explore further how hard the SNAILs can be driven with one or more drive tones.  This is directly related to the issue of saturation power in parametric amplifiers, where recent exciting results \cite{ sivak_kerr-free_2019, parker2021near} provide guidance on how we may further optimize our router. With stronger module-router couplings, it is feasible to push our overall gate time down to $\unit[\sim10]{ns}$ in better optimized, next generation devices.

\section*{Methods}
\textbf{Device fabrication.}
The device in Fig.~\ref{fig:schematic}d contains a SNAIL on a sapphire chip, three transmon qubits on individual sapphire chips, and multiple 3D resonator modes coupled with each other as shown in Fig.~\ref{fig:schematic}c.  The coupling between the SNAIL and waveguide modes is determined by the shape of the SNAIL antenna (Supplementary Figure 2b), which is fabricated using photolithography and acid etching from a 200-nm thin tantalum film on a c-plane sapphire substrate\cite{place2021new}. Two windows were opened on the 3D waveguide above and below the SNAIL chip in order to place a copper magnet and a pump port into the waveguide to enable flux bias and strong pumping \change{(Supplementary Figure 2a)}. The antenna of the transmon qubits are fabricated using the same tantalum etching technique, and both the SNAIL and transmon junctions are composed of Al-$\text{AlO}_x$-Al layers fabricated using a standard Dolan bridge method.

\textbf{SNAIL mode characterization.} The SNAIL mode is characterized by measuring the transmission signal from the SNAIL pump port to a side port on the waveguide using a network analyzer. By sweeping the bias current applied to the magnet, we can measure how the frequency of the SNAIL and waveguide modes are changed by flux biasing the SNAIL loop \change{(Supplementary Figure 2d)}.

\textbf{Experiment setup.}
The full device is installed at the base ($\unit[\sim18]{mK}$) plate of a cryogenic set-up \change{(Supplementary Figure 3)}. Here, all pulse sequences are generated by a Keysight M3202A (1~GSa/s) and M3201A (500~MSa/s) Arbitrary Waveform Generators (AWGs). The baseband microwave control pulses are generated at an intermediate frequency (IF) of 100 MHz and upconverted to microwave frequencies using IQ mixers. Image rejection (IR) mixers have been used for downconverting the detected signals to 50 MHz, which are then digitized using a control system based on Keysight M3102A Analog-to-Digital converters with a sampling rate of 500 MSa/s and on-board Field-Programmable Gate Arrays (FPGA) for signal processing.

\textbf{Numerical simulations.}
We simulate the behavior of our system by analyzing the behavior of seven modes participating in the experiments: three qubit modes and four communication modes.  We treat the gates as ideal parametric interactions, and work in the rotating frame of the system (\change{details in Supplementary Method 1}). The Hamiltonian then contains single-qubit controls, cavity-cavity inter-module interactions, and qubit-cavity intra-module interactions listed respectively to give:

\begin{widetext}
\begin{equation}
    \label{eq:qubitAndCavity}
    \hat{\mathcal{H}}_{QC}/\hbar = \sum_{m=2,3,4} \eta_m ({\hat{q}_m}^\dagger + {\hat{q}_m}) + \sum_{\substack{i,j=1,2,3,4\\ i \neq j}} \eta_{ij} \ g_{c_i c_j s} \left( {\hat{c}_i}^\dagger {\hat{c}_j} + {\hat{c}_i} {\hat{c}_j}^\dagger \right) + \sum_{k=2,3,4} \eta_{k} \ g_{q_k q_k c_k c_k} \left( {\hat{c}_k}^\dagger {\hat{q}_k} + {\hat{c}_k} {\hat{q}_k}^\dagger \right),
\end{equation}
\end{widetext} 
where $\hat{q}_i$ indicates the qubit mode in each module and $\eta(t)$ represents the time-dependent strength of a given pulse, which follows the shapes and durations used in the experiment. To capture the effects of photon loss and decoherence in the system, we add loss operators with rates corresponding to the measured values listed in SI Table~1 and simulate the evolution of the system via the Lindblad master equation\cite{breuer2002theory} using QuTiP\cite{johansson_qutip_2013}:

\begin{equation}
    \dot{\rho}(t) = -\frac {i}{\hbar} \left[\hat{H}_{QC} (t), \rho (t) \right] + \sum_n \mathcal{D}\left[\hat{C}_n\right] (\rho)
\end{equation}
where $\rho$ represents the density matrix of the system and $\mathcal{D}[C_n]\left(\rho\right) = \hat{C}_n \rho  \hat{C}_n^\dagger - 1/2 \left(\hat{C}_n^\dagger \hat{C}_n \rho + \rho\hat{C}_n^\dagger \hat{C}_n \right)$ is the interaction between the system and the environment for different collapse operators.

For all experiments reported in the main text, we apply identical pulse sequences in the simulations, and record the final states after half of the measurement time (to account for decay during the measurement process). Results are consistent between experiments and simulation, which indicates that our device is primarily limited by the coherence times of our modes.

\textbf{Data processing.} 
For all quoted fidelities in the main text, we have first reconstructed the density matrix $\rho$ from tomographic measurements, and for a given target state $\sigma$, the fidelity of the results is calculated using:

\begin{equation}
F(\rho, \sigma) = \left(\text{tr} \sqrt{\sqrt{\rho} \sigma\sqrt{\rho}}\right)^2.
\end{equation}

Furthermore, a bootstrap method\cite{young_everything_2014} has been used to estimate the uncertainty of the reported fidelity. In experiments, all final datasets contain more than 10,000 averages;  we restructure the data set into $N_{\text{boot}}=1,000$ data sets each containing $N=10,000$ points obtained by Monte Carlo sampling of the original set of $10,000$ points. During Monte Carlo sampling, the probability that a data point is picked is 1/N irrespective of whether it has been picked before. In the end, we calculate the standard deviation of the bootstrap data sets $s_{x^B}$.

In general, $s_{x^B}$ should be related to the uncertainty of the original sample $\sigma_x$ by:

\begin{equation}
    \sigma_x = \sqrt{\frac{N}{N-1}} s_{x^B}
\end{equation}

Since, in our case, $N=10,000$ is sufficiently large, we have $\sigma_x \approx s_{x^B}$.

\section*{Data availability}
The data and code that support the findings of this study are available from the corresponding authors upon reasonable request.

\section*{Acknowledgements}
The authors gratefully acknowledge the skilled machining and advice of William Strang, and Shyam Shankar, Kevin Chou, Konrad Lehnert, Alex Jones, Evan McKinney, and Robert Schoelkopf for fruitful discussions. We also acknowledge Keysight Technologies, especially Kevin Nguyen, for helping make the three-module portions of this experiment possible, and Alex Place and Andrew Houck for help in developing tantalum-based transmon and SNAIL fabrication. This material is based upon work supported by the Air Force Office of Scientific Research under award number FA9550-15-1-0015.  This work as also partially supported by the Charles E. Kaufman Foundation of the Pittsburgh Foundation, as well as the Army Research Office under contracts W911NF-18-1-0144 and W911NF-15-1-0397.  

\section*{Author contributions}
M.H., D.P. and R.M. conceived the device structure and designed the experiment. C.Z., M.P., P.L. and M.X. simulated and designed the device. C.Z. and P.L. fabricated the device with help from T.C. and X.C.. C.Z. and P.L. developed measurement software, performed the experiments, analyzed the data, and carried out numeric simulation of the experiment result. R.K. contributed to measurement hardware setup and writing instrument control software. W.P. helped with 4-wave-mixing discussion, experiment design, and supervised the software development. M.H., P.L. and C.Z. wrote the manuscript. All authors discussed the results and commented on the manuscript. M.H. supervised the experiment. \change{C.Z and P.L. are co-first authors and contributed equally to this publication.}

\section*{Competing interests}
Authors Hatridge, Pekker, and Mong are co-inventors of the University of Pittsburgh patent application “Parametrically-driven coherent signal router for quantum computing and related methods” (Application number 17/686,702)  which covers the parametric controls and modular methods shown in this paper.

\bibliography{refs}
\bibliographystyle{naturemag}

\end{document}


\preprint{AIP/123-QED}

\title[]{Supplementary Information for `Realizing all-to-all couplings among detachable quantum modules using a microwave quantum state router'}

\makeatletter
\let\@fnsymbol\@fnsymbol@latex
\@booleanfalse\altaffilletter@sw
\makeatother

\author{Chao Zhou$^{1\dagger}$, Pinlei Lu$^{1\dagger}$, Matthieu Praquin$^2$, Tzu-Chiao Chien$^1$, Ryan Kaufman$^1$, Xi Cao$^1$, Mingkang Xia$^1$, Roger S.~K.~Mong$^1$, Wolfgang Pfaff$^3$, David Pekker$^1$, Michael Hatridge$^1$}

\affiliation{$^1$Department of Physics and Astronomy, University of Pittsburgh, Pittsburgh, PA, USA \\ 
$^2$Département de Physique, École Normale Supérieure, Paris, France \\ 
$^3$Department of Physics, University of Illinois Urbana-Champaign, Champaign, IL, USA \\ 
$^{\dagger}$These authors contributed equally to this publication.}

\date{\today}

\maketitle

\tableofcontents

\begin{center}
    \rule{10cm}{1pt}
\end{center}

\newpage
\section{Supplementary Figures}
\subsection{Detailed device schematic}
\begin{figure*}[h]
\includegraphics[]{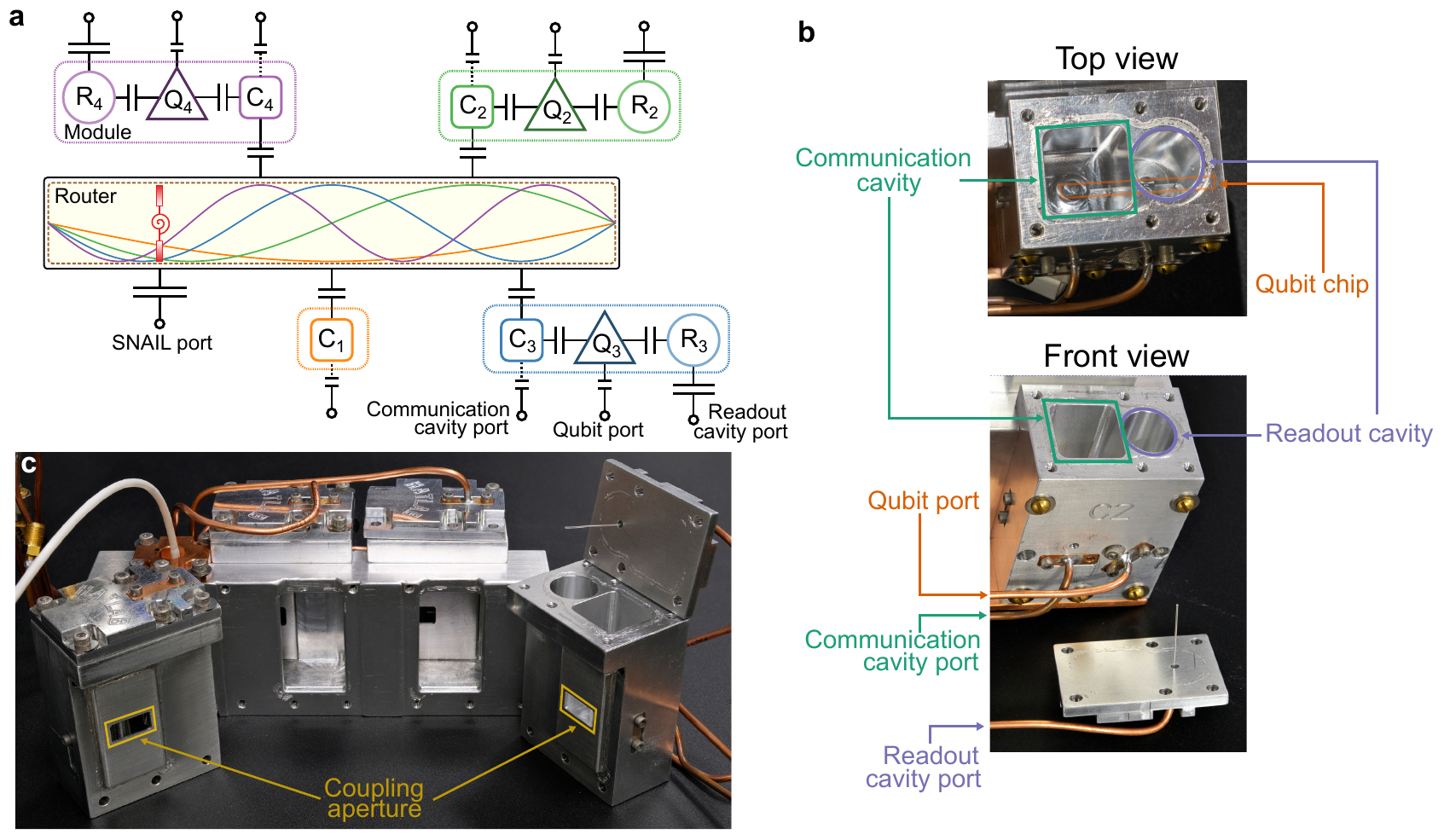}
\caption{\textbf{Schematic and images of our full device} (a) Schematic of the full device. Each communication cavity ($\textrm{C}_1 - \textrm{C}_4$) is coupled to one waveguide mode and a weak external port for initial characterization.  This weak port was omitted in the full experiment. Modules 2-4 each contain a qubit  ($\textrm{Q}_2 - \textrm{Q}_4$) that couples to both the communication mode and a readout cavity mode ($\textrm{R}_2 - \textrm{R}_4$). Each qubit and readout cavity also has a dedicated drive/readout port. (b) Top and front view of an open module. The communication cavity and readout cavity are both coaxial $\lambda/4$ post cavities with a hole drilled in their shared sidewall to accomodate the qubit.  A narrow, rectangular electrical-discharge-machine cut channel through the entire module allows the chip to slide in from the side; the qubit chip passes fully through both cavities. The external port for each mode is labeled in the front view picture. (c) Picture of the device partially disassembled. The sidewalls of the communication cavities serve as parts of the waveguide sidewall; the two sides are joined with an indium wire. The small apertures (yellow boxes) on these sidewalls allow electromagnetic field overlap between the waveguide modes and the cavity modes, thereby create the $g_{c \textrm{w}}$ couplings.}
\label{fig:coupling_scheme}
\end{figure*}

The full coupling scheme between modes and external ports in our device is shown Supplementary Figure \ref{fig:coupling_scheme}a.  Supplementary Figure \ref{fig:coupling_scheme}b shows the inside of a single qubit module consisting of three modes (the round readout post cavity, the square communication cavity, and the module qubit which spans the two cavities via a hole in their shared wall. The communication cavity, the qubit chip, the readout cavity and the external ports are labeled. Supplementary Figure \ref{fig:coupling_scheme}c shows a picture of the full device partially disassembled.

\newpage
\subsection{Details of the SNAIL mode}
\begin{figure*}[h]
\includegraphics[]{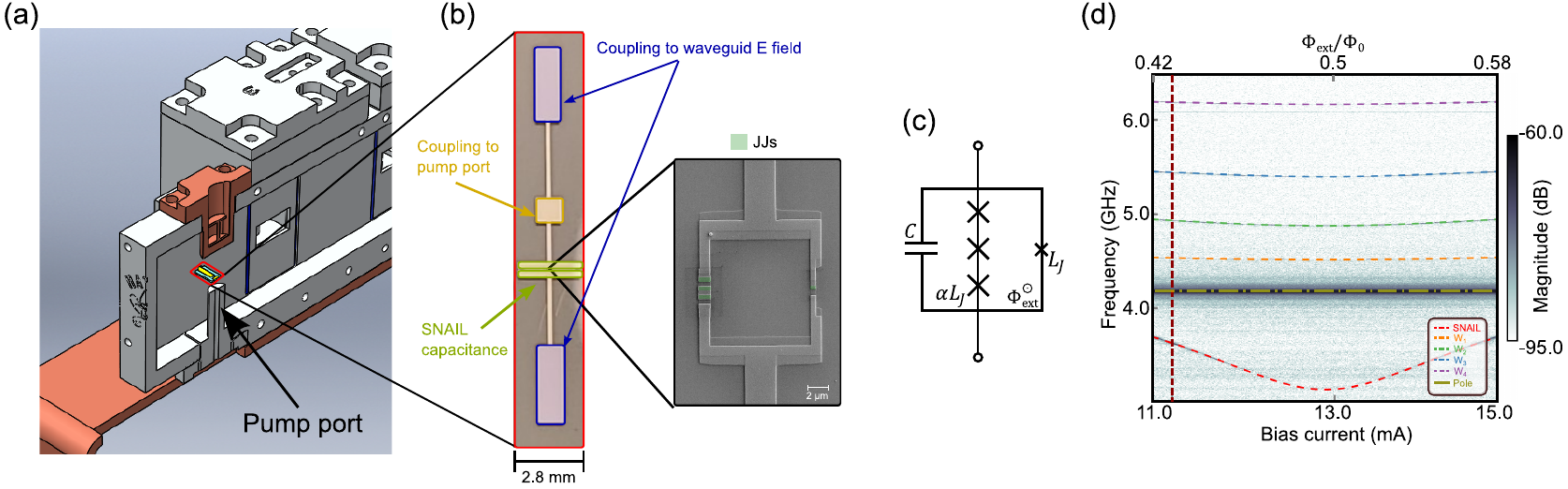}
\caption{\textbf{SNAIL mode details.} (a) Physical position of the SNAIL chip and pumping port.  The SNAIL is flux-biased via a copper-sheathed electromagnet protruding into the waveguide above it, and is strongly coupled to a microwave drive line introduced via an aluminum cylinder below the SNAIL. (b) False color optical and SEM image of the SNAIL, indicating its essential components. (c) Circuit diagram of the SNAIL device. The device consist of a superconducting loop of a small Josephson junction (with Josephson inductance $L_J$) in parallel with three identical larger junctions, each has Josephson inductance of $\alpha L_J$. The junction loop is also in parallel with a capacitance $C$ and is biased with external flux $\Phi_{\text{ext}}$ (d) Color plot of the magnitude of transmission signal versus the frequency from SNAIL pumping port ($|S_{21}(\omega)|$) for a range of applied coil bias current/applied SNAIL flux. The dotted lines indicate the dressed modes of the waveguide modes and SNAIL, as well as the `pole mode' of the aluminum cylinder containing the drive port.  The vertical, dark-red dashed line indicates the operating flux of the SNAIL.}
\label{fig:fluxSweep}
\end{figure*}

\newpage
\subsection{Experimental wiring diagram}
\begin{figure*}[h]
\includegraphics[]{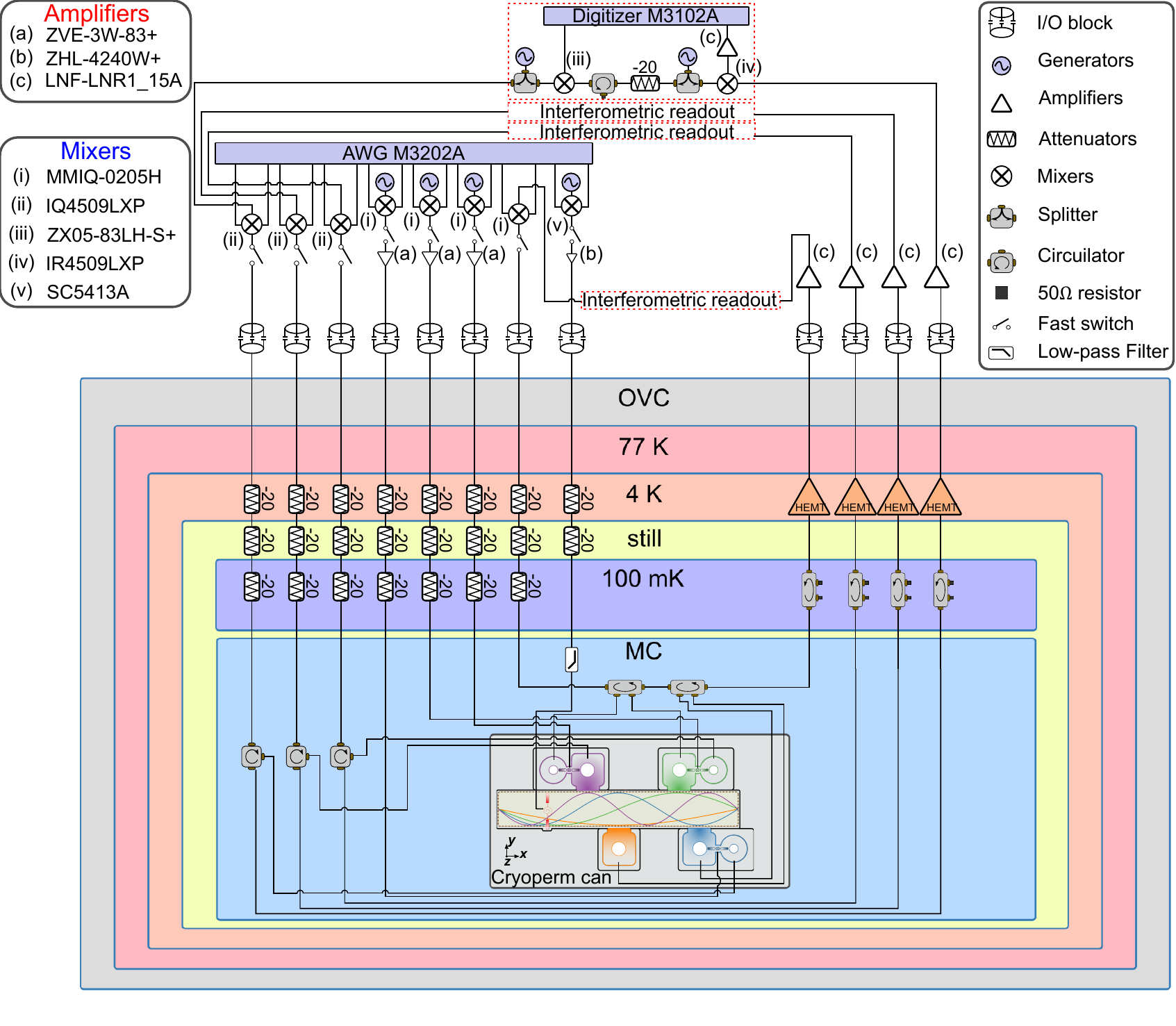}
\caption{\textbf{Experimental wiring diagram} }
\label{fig:wiring}
\end{figure*}

\newpage
\section{Supplementary Tables}
\subsection{Mode parameters and frequency spectrum}
Supplementary Table \ref{table:allmodes} contains the measured parameters of all key modes in our device. All values have been directly measured and calibrated in the experiments at the operation bias point of the SNAIL. They are also used in our master equation simulations (see Methods, Numerical simulations in the main text). The measurement fidelity is calculated using single-qubit tomography. The single-qubit gate fidelity is measured using interleaved randomized benchmarking\cite{kelly2014optimal}. In previous experiments we measured the router's waveguide mode frequencies and lifetimes, which we summarize in Supplementary Table \ref{table:wgmodes}.  For qubits the coherence times $T_1$, $T_{2R}$ and $T_{2E}$ are the energy decay lifetime, coherence time measured via a Ramsey sequence, and coherence time measured via a Hahn Echo sequence, respectively.  For cavities the coherence times $T_1$ and $T_2$ are measured by either (a) using an under-coupled probe port to drive a coherence state and measure the voltage decay or (b) swapping single photons from qubits to cavities for variable time.  The two times are listed as  a (b) in the table.  The pure dephasing time $T_{\phi}$ can be inferred from these values using $T_{2}=1/[1/(2T_{1}) + 1/T_{\phi}]$.

The SNAIL mode lifetime is primarily limited by coupling to the copper intrusion of the bias magnet and coupling to the drive port, which thereby also limits the lifetime of the waveguide modes. The loss from the drive port is reduced in the full experiment by the low-pass filter on the pump line.  However, we did not directly measure waveguide mode lifetimes as the same filter renders them un-measurable via any port. With the filter added, we believe the major loss sources for the waveguide modes are the seam losses at joints between the waveguide and the communication cavity modes, and the loss due to the copper intrusion.

\begin{table}[h]
\begin{tabular}{|l|l|l|l|l|l|l|l|l|}
\hline
                       & $Q_2$               & $Q_3$              & $Q_4$              & $C_1$            & $C_2$      & $C_3$      & $C_4$      & SNAIL \\ \hline
$\omega_0/2\pi$ (GHz)    & 3.067984            & 4.040709           & 3.566572           & 4.477662         & 4.812500   & 5.474195   & 6.180769   & 3.914900                  \\ \hline
$\text{T}_1 \ (\mu s)$ & 60              & 9.1              & 8.4              & 22 (23) & 23 (27)     & 13 (13)     & 15 (20)     & 0.98                     \\ \hline
$\text{T}_\text{2R} \ (\mu s)$& 18       & 6.3              & 8.0              & 44 (45)   & 46 (47)     & 22 (11)     & 25 (23)     & 1.0                   \\ \hline
$\text{T}_\text{2E} \  (\mu s)$& 24     & 7.6              & 8.0              &                  &            &            &            &                           \\ \hline
$\alpha/2\pi$ (MHz)    & -141.3               & -118.1              & -125.8              &                  &            &            &            &                           \\ \hline
$\chi_{qc} / 2\pi$ (MHz)    & -0.11               & -1.7              & -0.86              &                  &            &            &            &                           \\ \hline
Measurement fidelity & 93.6\%             & 83.0\%            & 88.0 \%           &                  &            &            &            &                           \\ \hline
Single gate error   & 0.48\% $\pm$ 0.04\% & 3.74\% $\pm$ 0.5\% & 3.21\% $\pm$ 0.6\% &                  &            &            &            &                           \\ \hline
\end{tabular}
\\
\caption{\textbf{Devices parameters.}}
\label{table:allmodes}
\end{table}

\begin{table}[h]
\begin{tabular}{|l|l|l|l|l|}
\hline
                       & $W_1$ & $W_2$ & $W_3$ & $W_4$ \\ \hline
$\omega_0/2\pi$ (GHz)  & 4.534 & 4.936 & 5.446 & 6.190 \\ \hline
$\text{T}_1 \ (\mu s)$ & 1.68  & 0.29  & 0.28  & 0.81  \\ \hline
\end{tabular}
\caption{\textbf{Waveguide mode lifetimes.}}
\label{table:wgmodes}
\end{table}

\begin{figure*}[h]
\includegraphics[]
{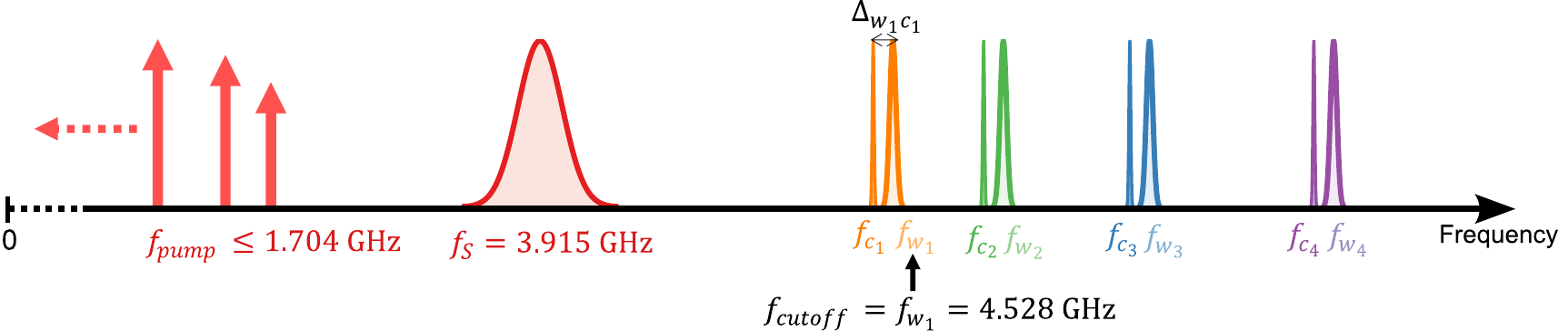}
\caption{\textbf{Frequency spectrum of all linear modes, SNAIL mode, and pumping frequencies.} The maximum pumping frequency on the SNAIL is $f_{pump}^{max} \approx f_{c_4}-f_{c_1} = 1.704 \ \text{GHz}$, which is far below the waveguide cutoff frequency $f_{cutoff} = f_{\text{w}1} = 4.528 \ \text{GHz}$. The router then naturally protects pumping tones from propagating into modules. Meanwhile, the frequencies of all other modes don't have to be precisely controlled, so the router can be easily adapted to different modules without fine-tuning. Since the maximum pumping frequency is also lower than the SNAIL frequency $f_{pump}^{max} < f_s = 3.915~\text{GHz}$, a low-pass filter (e.g. Mini-Circuits VLF-2250+) is added to the SNAIL pump port so that strong pumps can be applied to SNAIL while the SNAIL mode lifetimes are protected.}
\label{fig:frequencySpectrum}
\end{figure*}

One natural advantage of our router design is that the parametric pumps are well-separated in frequency from all modes. This allows us to protect the quantum modes while still strongly pumping the parametric processes. As we can see from Supplementary Figure \ref{fig:frequencySpectrum}, the communication modes and SNAIL mode itself are all at least \unit[2]{GHz} higher than the parametric pumping frequencies.

\newpage
\subsection{$\sg$ gate time for all six cavity pairs}
By preparing coherent states in all four different cavities, we can perform six pairs of inter-module $\sg$s. We list all \sg~ gate times in Supplementary Table~\ref{tab:swapTimes}.  The average gate time is $\unit[764]{ns}$.
\begin{table}[htbp!]
\begin{tabular}{|c|r|}
\hline
$\sg$ pair                   & $\sg$ time (ns)  \\ \hline
$C_1 \leftrightarrow C_2$   & 1248            \\ \hline
$C_1 \leftrightarrow C_3$   & 651             \\ \hline
$C_1 \leftrightarrow C_4$   & 535             \\ \hline
$C_2 \leftrightarrow C_3$   & 942             \\ \hline
$C_2 \leftrightarrow C_4$   & 832             \\ \hline
$C_3 \leftrightarrow C_4$   & 375             \\ \hline
\end{tabular}
\caption{\textbf{$\sg$ gate times between all six possible communication cavity pairs.} The data is measured by preparing a coherent state in one cavity and swapping it to another cavity. For each swap pair, the listed $\sg$ gate time is based on the maximum $\sg$ speed measured by tuning up the SNAIL pump power until we see obvious decrease of mode lifetime.}
\label{tab:swapTimes}
\end{table}

\newpage
\section{Supplementary Methods}

\subsection{Theory of Parametric photon $\sg$ gate using 3-wave-mixing}
\label{s_sec: thoery_3wm}
Here we demonstrate the theory of 3-wave-mixing parametric exchange process in the system. Without loss of generality, we have excluded the qubit and readout cavity modes from the total Hamiltonian, and focused on the parametric coupling between the communication cavities.
\begin{equation}
\label{eq:RandRC}
\begin{split}
    \Hamiltonian_0 / \hbar = & \sum_{i=1,2,3,4} \left[\omega_{\textrm{w}_i} \linmode{\textrm{w}_i}{} +  g_{\textrm{w}_{i}s} (\hat{\textrm{w}_i}^\dagger \hat{s} + \hat{\textrm{w}_i} \hat{s}^\dagger)\right] + \sum_{i=1,2,3,4}  \left[ \omega_{c_i} \linmode{c}{i} + g_{c_i \textrm{w}_i} (\hat{c_i}^\dagger \hat{\textrm{w}_i} + \hat{c_i} \hat{\textrm{w}_i}^\dagger)\right]\\
    & \ + \omega_s \linmode{s}{} + g_3 (\hat{s} + \hat{s}^\dagger)^3
\end{split}
\end{equation}
After re-diagonalizing the second-order terms in Eq.~\ref{eq:RandRC}, all dressed modes in the router inherit third-order nonlinearity. The terms we use for the parametric pumping scheme are three-wave mixing terms:
\begin{equation}
\label{eq:3waveMixingTerm}
g_{c_i c_j s} \left(\hat{c_i}^\dagger \hat{c_j} \hat{s} + \hat{c_i} \hat{c_j}^\dagger \hat{s}^\dagger\right), \  (i, j = 1, 2, 3, 4; i\ne j).    
\end{equation}
Here, the $\hat{c_i} \text{~and~} \hat{s}$ operators represent the dressed modes after re-diagonalization, and $g_{c_i c_j s}$ is the 3-wave-mixing coefficient. In the weak coupling regime where $(\frac{g}{\Delta})_{c_i \textrm{w}_i}, (\frac{g}{\Delta})_{\textrm{w}_is} \sim 0.1$, the coefficient ${g_{c_i c_j s} \approx g_3 (\frac{g}{\Delta})_{c_i \textrm{w}_i} (\frac{g}{\Delta})_{\textrm{w}_is} (\frac{g}{\Delta})_{c_j w_j} (\frac{g}{\Delta})_{w_js}}$, where $\Delta_{\textrm{w}_i s} = \omega_{\textrm{w}_i} - \omega_s$.

To realize $\sg$ gates between communication cavities based on Eq.~\ref{eq:3waveMixingTerm}, we apply a strong single-tone pump on the SNAIL at the frequency difference of two cavities, $\omega_p = \left| \omega_{c_i} - \omega_{c_j} \right|$, where $i,j = 1, 2, 3, 4 ; i\ne j$. The time-dependent pumping term can be written as $ \hat{\mathcal{H}}_{P} / \hbar = \epsilon(t) ( \hat{s}+\hat{s}^\dagger)$, where $\epsilon(t)$ is represented by:
\begin{equation}
    \epsilon(t) = 
    \begin{cases}
        \epsilon^x (t) \cos (\omega_p t) + \epsilon^y (t) \sin(\omega_p t),   & 0 < t < t_g \\
        0, & \text{otherwise}.
    \end{cases}
\end{equation}
Thus, the total system Hamiltonian under pumping can be written as $\Hamiltonian' = \Hamiltonian_0 + \Hamiltonian_P$. To study the effect of this pumping term in the total Hamiltonian, we apply a unitary transformation on $\Hamiltonian$ with displacement operator:
\begin{equation}
    D(t) = \exp [(z s^\dagger - z^* s)],
\end{equation}
where $z = -\frac{(\epsilon^x + i \epsilon^y)/2}{\omega_p - \omega_s}e^{-i \omega_p t} + \frac{(\epsilon^x - i \epsilon^y)/2}{\omega_p + \omega_s} e^{i\omega_p t}$. This results in a new Hamiltonian $\hat{\mathcal{H}}^D$, in which the $ \hat{\mathcal{H}}_{P}$ term is canceled and $\hat{s} \rightarrow \hat{s}-z$. Specifically, the 3-wave mixing term in Eq.~\ref{eq:3waveMixingTerm} becomes:

\begin{equation}
\label{eq:drive3Wave}
g_{c_i c_j s} \hat{c_i}^\dagger \hat{c_j} \left(\hat{s} \frac{(\epsilon^x + i \epsilon^y)/2}{\omega_p - \omega_s}e^{-i \omega_p t} - \frac{(\epsilon^x - i \epsilon^y)/2}{\omega_p + \omega_s} e^{i\omega_p t} \right) \\ + h.c.   
\end{equation}

Then we apply a rotating frame transformation at the frequency of all the router modes (SNAIL + waveguide) and communication modes in the system:
\begin{equation}
\label{eq:RFTOperator}
    R(t) = \exp \left[i\omega_s \hat{s}^\dagger\hat{s} +\sum_{i=1,2,3,4} i\left( \omega_{\textrm{w}_i}\hat{\textrm{w}_i}^\dagger\hat{\textrm{w}_i} + \omega_{c_i}\hat{c_i}^\dagger\hat{c_i} \right)\right],
\end{equation}
Note that $\omega_p = \left| \omega_{c_i} - \omega_{c_j} \right|$. We assume that all the other frequency differences in the system are several linewidths away from the pumping frequency. Then the only slowly rotating term after the transformation is

\begin{equation}
\label{eq:H_RWA}
    \Hamiltonian^{\text{RWA}}/\hbar= \eta \ g_{c_i c_j s} \left( {\hat{c}_i}^\dagger {\hat{c}_j} + {\hat{c}_i} {\hat{c}_j}^\dagger \right),
\end{equation}
where $\eta = (\epsilon^x + i \epsilon^y) \omega_s/(\omega_d^2 - \omega_s^2) $. Here, $\eta$ corresponds to the square root of the effective pump photon number in the SNAIL mode, i.e.  $\eta \equiv \sqrt{n_s}$.

Equation~\ref{eq:H_RWA} shows that pumping at $\omega_p = \left| \omega_{c_i} - \omega_{c_j} \right|$ activates the spontaneous photon exchange process between communication mode $c_i$ and $c_j$. The $\sg$ speed depends on the pumping strength and the three-wave mixing coefficient $g_{c_ic_js}$. By controlling the length of the pulse, $\sg^{1/n}$ gates can be performed between the two modes. Also, the existence of all the 3-wave-mixing terms between different modes allows us to pump multiple parametric processes simultaneously, e.g. we can perform $\sg$ gate between two different cavity pairs at the same time by simultaneously pumping the SNAIL mode at the two difference frequencies.

\subsection{Tune up intra-module $\sg$ interaction}
The intra-module $\sg$\cite{narla2016robust} interaction swaps information between qubits and communication cavities. In the experiment, the qubit is first prepared to its excited state, which gives module state $\ket{e0}$. Next, two side-band tones are applied near to the qubit and cavity frequencies. Here the detunings are around 10$\kappa$ ($\sim 20$ MHz) away from each mode to avoid any undesired mode excitation, as shown in Supplementary Figure \ref{fig:intraModuleSWAP}. Then, both $\sg$ time and cavity side-band (CSB) detuning $\delta$ are swept to determine the optimized $\sg$ condition. The average fidelity of a $\sg$ gate is $94 \%$, which is mainly limited by how fast the $\sg$ gate is compared to $T_2$ decoherence of qubit and cavity.

\begin{center}
\begin{figure*}[h]
\centering
\makebox[\textwidth]{\includegraphics{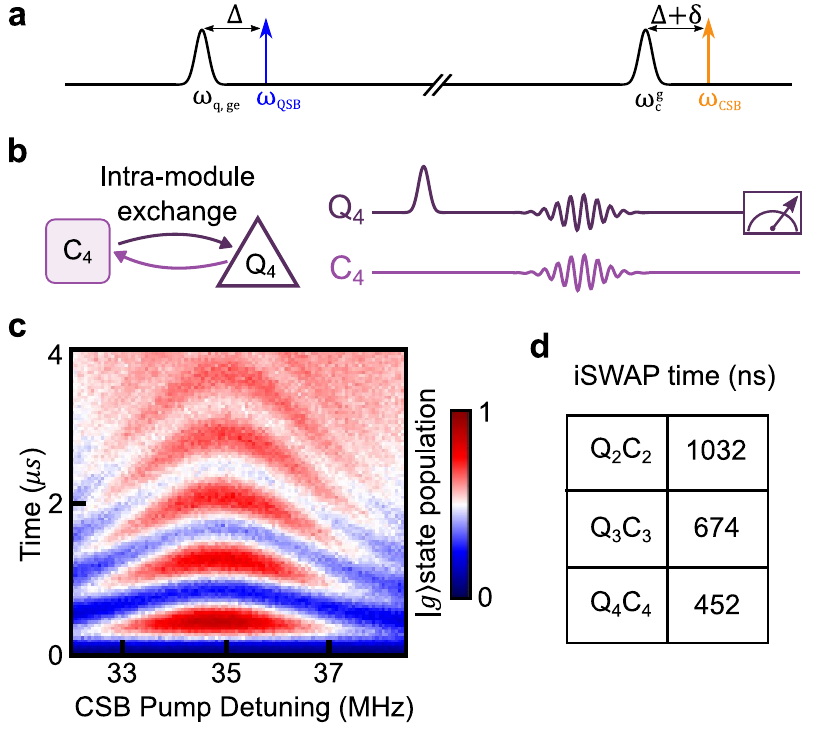}}
\caption{\textbf{Intra-module $\sg$.} (a) Schematic of qubit and communication cavity frequencies and pump tones. Two pump tones have been applied near the qubit and cavity frequencies to generate exchange interaction between qubit and cavity. (b) Schematic and pulse sequence. The qubit is prepared in its excited state via a $R_x(\pi)$ pulse at $\omega_{q, ge}$, then two sideband pulses are applied at $\omega_{\text{QSB}}$, detuned by $\Delta$ from $\omega_{q, ge}$, and $\omega_{\text{CSB}}$, detuned by $\Delta + \delta$ from $\omega_c^g$. Finally, projective measurement is applied to the qubit. Both pulse length, and $\delta$ are swept to determine an optimized $\sg$ gate. (c) Rabi-like oscillation of the intra-module $\sg$ interaction. The vertical axis is the pulse time, and the horizontal axis is the CSB detuning $\delta$, where the color indicates the $\ket{g}$ state population of qubit. (d) Full $\sg$ time for all three cavity-qubit pairs in the system}
\label{fig:intraModuleSWAP}
\end{figure*}
\end{center}

\newpage
\section{Supplementary Discussion}
\subsection{Exchange interaction leakage mitigation}
Our router design contains intermediate waveguide modes which are close in frequency to the communication cavity modes. Thus, it is possible that when we pump an exchange process between two cavities strongly, their states could leak into to the neighbouring waveguide modes. Here we show how this leakage can be fixed using the idea of ``derivative removal by adiabatic gate (DRAG)" \cite{motzoi_simple_2009} at least for moderately long-lived waveguide modes.

Here we only consider the exchange process between two cavity modes $c_{1} \text{~and~} c_{2}$ with their possible leakage to waveguide modes $\textrm{w}_{1} \text{~and~} \textrm{w}_{2}$. Following the same derivations as we have done in Supplementary Method \ref{s_sec: thoery_3wm} Eq.~\ref{eq:RandRC}-\ref{eq:drive3Wave}, to see the effect of the leakage to waveguide mode, we write the rotating frame transformation operator (Eq.~\ref{eq:RFTOperator}) in a slightly different way:
\begin{equation}
    R'(t)  = \exp{\left[ i\omega_{s}  \hat{s}^\dagger \hat{s} + i(\omega_{c_1} - \delta/2) \hat{c}_1^\dagger \hat{c}_1 + i(\omega_{c_2} + \delta/2) \hat{c}_2^\dagger \hat{c}_2 + i(\omega_{c_1} - \delta/2) \hat{\textrm{w}}_1^\dagger \hat{\textrm{w}}_1 + i(\omega_{c_2} + \delta/2) \hat{\textrm{w}}_2^\dagger \hat{\textrm{w}}_2 \right]},
\end{equation}
where $\delta = \omega_{c_1} - \omega_{c_2} - \omega_p $. Then the Hamiltonian after RWA becomes:
\begin{equation}
\label{eq:H_RWAWithWG}
\begin{split}
    \Hamiltonian'^{\text{RWA}}/\hbar = & \frac{\delta}2 (\hat{c}_1^\dagger \hat{c}_1 - \hat{c}_2^\dagger \hat{c}_2) + (\delta/2 + \Delta_1) \hat{\textrm{w}}_1^\dagger \hat{\textrm{w}}_1 + (-\delta/2 + \Delta_2) \hat{\textrm{w}}_2^\dagger \hat{\textrm{w}}_2 \\
     & + \eta(g_{c_1 c_2 s} \hat{c}_1^\dagger \hat{c}_2 + g_{c_1 \textrm{w}_2 s} \hat{c}_1^\dagger \hat{\textrm{w}}_2 + g_{c_2 \textrm{w}_1 s} \hat{c}_2^\dagger \hat{\textrm{w}}_1)
     + \text{h.c.},
\end{split}
\end{equation}
where $\Delta_i = \omega_{\textrm{w}_i} - \omega_{c_i}$. Here, the cavity-cavity exchange term that we have seen in Eq.~\ref{eq:H_RWA} still  is present in Eq.~\ref{eq:H_RWAWithWG}. However, for $\eta g_{c_i w_j} \sim \delta+\Delta_j$ (strong pump), the effects of $\hat{c}_i^\dagger \hat{\textrm{w}}_j$ terms (waveguide-cavity exchange) can no longer be neglected. To study how this effect can be canceled using DRAG method, we introduce the adiabatic transformation V that allows us to work entirely in the cavity-cavity subspace. This transformation is 
\begin{equation}
    V(t) = \exp{ \left[ -i \Re(\eta) \left(i \left(\frac{g_{c_1 \textrm{w}_2 s}}{\Delta_2 - \delta} \hat{c}_1 \hat{\textrm{w}}_2^\dagger + \frac{g_{c_2 \textrm{w}_1 s}}{\Delta_1 + \delta} \hat{c}_2 \hat{\textrm{w}}_1^\dagger\right) + \text{h.c.} \right)  \right] }.
\end{equation}
After the adiabatic transformation, we have 
\begin{equation}
    \hat{c}_i \rightarrow \hat{c}_i + \Re(\eta) \zeta_i \hat{\textrm{w}}_j, ~~
    \hat{\textrm{w}}_j \rightarrow \hat{\textrm{w}}_j - \Re(\eta) \zeta_i \hat{c}_i
\end{equation}
where $i,j=1,2;~ i\neq j; ~\zeta_1=\frac{g_{c_1 \textrm{w}_2 s}}{\Delta_2-\delta}; ~ \zeta_2=\frac{g_{c_2 \textrm{w}_1 s}}{\Delta_1+\delta}.$ After the transformation, the leakage terms: $\hat{c}_1^\dagger \hat{\textrm{w}}_2 $ and $ \hat{c}_2^\dagger \hat{\textrm{w}}_1$ become:
\begin{equation}
\begin{split}
    \frac{i g_{c_1 \textrm{w}_2 s}\omega_s}{\omega_d^2 - \omega_s^2}\left( \epsilon^y +  \frac{\dot{\epsilon^x}}{\Delta_2-\delta} \right) \hat{c}_1^\dagger \hat{\textrm{w}}_2 + \frac{i g_{c_2 \textrm{w}_1 s}\omega_s}{\omega_d^2 - \omega_s^2}\left( \epsilon^y +  \frac{\dot{\epsilon^x}}{\Delta_1 +\delta} \right) \hat{c}_2^\dagger \hat{\textrm{w}}_1 + h.c.,
\end{split}
\end{equation}
We find these terms can be cancelled under the condition:
\begin{equation}
    \epsilon^y = - \frac{\dot{\epsilon^x}}{\Delta_2-\delta} \ \text{and} \ \epsilon^y = - \frac{\dot{\epsilon^x}}{\Delta_1+\delta}.
\end{equation}

Also, ac-Start shift (phase) error can be eliminated with the detuning $\delta$ that satisfies:
\begin{equation}
    \delta/2 + \Re^2{(\eta)} \frac{g_{c_1 \textrm{w}_2 s}^2}{(\Delta_2-\delta)^2}\left( -\Delta_2 + \frac{3}{2} \delta \right) =0 ;~ \delta/2 + \Re^2{(\eta)} \frac{g_{c_2 \textrm{w}_1 s}^2}{(\Delta_1+\delta)^2}\left( \Delta_1 + \frac{3}{2} \delta \right) =0 
\end{equation}

Thus, we have eliminated the leakage to the nearest waveguide mode, however, we also notice that this frame induces the term $c_i^\dagger \textrm{w}_i$ with $\eta^2$. This error can also be removed by adding higher transformation to the system, but because of the fact that we drive the system much detuned, this term can be ignored when $\eta << 1$.

\newpage
\subsection{Effect of exchange interaction VS mode type and state encoding}
As noted in the main text, depending on the qubit encoding in the Hilbert space of the state being swapped, the same parametric coupling (i.e. $\hat{\mathcal{H}}^{\text{eff}} = g^\text{eff}_{c_i c_j} (\chat_i^\dagger \chat_j + \chat_i \chat_j^\dagger )$) can result in very different state evolution. To clarify this, we compare different initial states and different kinds of modes (transmon qubit or cavity) at times ${t = \pi/g^\text{eff}, 2\pi/g^\text{eff}, 3\pi/g^\text{eff} \text{~and~} 4\pi/g^\text{eff}}$. In the following tables, the column labels indicate initial states, row labels indicate the unitary applied to the state, and the table values indicate the new state produced.

As shown in Supplementary  Table \ref{tab:tlsTotls}, the effective conversion process can perform a perfect \sg ~ gate between two qubits. However, if the coupled objects include one linear mode, the situation becomes more complicated. For example, as shown in Supplementary  Table \ref{tab:cavTocav}, in the Fock state basis, the conversion process between two linear modes is more like a `beam splitter' process\cite{pfaff2017controlled, burkhart2020error}; and in the coherent state basis, the process becomes a non-entangling exchange.

\begin{table}[htbp!]
\begin{tabular}{|c!{\vrule width 2pt}c|c|c|}

\hline

\textbf{$\ket{\psi_{\text{init}}}$} & $\ket{1, 0}$ & $\ket{1, 1}$ & $\alpha\ket{0, 0}+\beta\ket{0, 1}+\gamma\ket{1, 0}+\eta\ket{1, 1}$ \\ \noalign{\hrule height 2pt}

\textbf{$U^{1/2}\ket{\psi_{\text{init}}}$} & $\frac{1}{\sqrt{2}} (\ket{1, 0} - i \ket{0, 1})$ & $\ket{1, 1}$ & $\alpha\ket{0, 0} + \frac{1}{\sqrt{2}}(\beta - i\gamma)\ket{0, 1} + \frac{1}{\sqrt{2}}(\gamma - i\beta)\ket{1, 0} + \eta\ket{1, 1}$ \\ \hline

\textbf{$U\ket{\psi_{\text{init}}}$}  & $-i\ket{0, 1}$ &  $\ket{1, 1}$  & $\alpha\ket{0, 0} -i\gamma\ket{0, 1} - i\beta\ket{1, 0} + \eta\ket{1, 1}$ \\ \hline

\textbf{$U^{3/2}\ket{\psi_{\text{init}}}$} & -$\frac{1}{\sqrt{2}} (\ket{1, 0} + i \ket{0, 1})$ & $\ket{1, 1}$ &  $\alpha\ket{0, 0} - \frac{1}{\sqrt{2}}(\beta + i\gamma)\ket{0, 1} - \frac{1}{\sqrt{2}}(\gamma + i\beta)\ket{1, 0} + \eta\ket{1, 1}$
\\ \hline

\textbf{$U^2\ket{\psi_{\text{init}}}$}  & $-\ket{1, 0}$ & $\ket{1, 1}$ & $\alpha\ket{0, 0}-\beta\ket{0, 1}-\gamma\ket{1, 0}+\eta\ket{1, 1}$   \\ \hline
\end{tabular}
\caption{\textbf{Transmon to transmon exchange.} We consider the final states of different initial states under evolution operation $U^p$, where $U$ is defined as $U=\exp{(-i\hat{\mathcal{H}}t)}$, as $t= 2\pi/g^\text{eff}$. The exponent of $U$ simply means $U^p=\exp{(-i~\hat{\mathcal{H}}(pt))}$. Here, $U^p$ creates an \sg ~ family gate between the two qubits;  $U^{1/2}=\sqrt{\text{\textit{i}SWAP}}$ ~ could be used to create a Bell state given an initial state $\ket{1, 0}$. Note that $U$ performs an \sg ~ gate, which is also a universal two-qubit gate that can create entanglement. }
\label{tab:tlsTotls}
\end{table}

\begin{table}[htbp!]
\begin{tabular}{|c!{\vrule width 2pt}c|c|}
\hline

\textbf{$\ket{\psi_{\text{init}}}$}  & $\ket{1, 0}$ & $\ket{1, 1}$ \\ \noalign{\hrule height 2pt}

\textbf{$U^{1/2}\ket{\psi_{\text{init}}}$} & $\frac{1}{\sqrt{2}} (\ket{1, 0} - i \ket{0, 1})$ & $\frac{1}{\sqrt{2}} (\ket{1, 1} - i\ket{0, 2})$ \\ \hline

\textbf{$U\ket{\psi_{\text{init}}}$}  & $-i\ket{0, 1}$ &  $-i \ket{0, 2}$ \\ \hline

\textbf{$U^{3/2}\ket{\psi_{\text{init}}}$} & -$\frac{1}{\sqrt{2}} (\ket{1, 0} + i \ket{0, 1})$ & $-\frac{1}{\sqrt{2}} (\ket{1, 1} + i\ket{0, 2})$ \\ \hline

\textbf{$U^2\ket{\psi_{\text{init}}}$}  & $-\ket{1, 0}$ & $-\ket{1, 1}$ \\ \hline
\end{tabular}
\\
\footnotesize{\hspace*{4mm} * Here, the unitary is defined as $U=\exp{(-i\hat{\mathcal{H}}t')}$, where $t'=t/\sqrt{2}$.}
\caption{\textbf{Transmon to cavity exchange.} We consider the Fock state basis as for transmon-cavity coupling, coherent states could result in an even more complicated scenario. Here, the $U^p$ operator can still behave like a tunable beam-splitter if either the cavity or qubit is prepared to an empty/ground state. However, for initial states like $\ket{1, n}$, where $n\ge1$, the $U^p$ operator will no longer keep the state unchanged.}
\label{tab:tlsTocav}
\end{table}

\begin{table}[htbp!]
\begin{tabular}{|c!{\vrule width 2pt}c|c|c|c|}
\hline

\textbf{$\ket{\psi_{\text{init}}}$}  & $\ket{1, 0}$ & $\ket{1, 1}$ & $\ket{\alpha, 0}$  & $\ket{\alpha, \beta}$ \\ \noalign{\hrule height 2pt}

\textbf{$U^{1/2}\ket{\psi_{\text{init}}}$} & $\frac{1}{\sqrt{2}} (\ket{1, 0} - i \ket{0, 1})$ & $\frac{1}{\sqrt{2}} (\ket{2, 0} + \ket{0, 2})$ & $\ket{\alpha/2, -i~\alpha/2}$ & $\ket{\alpha/2 - i~\beta/2, \beta/2 - i~\alpha/2}$ \\ \hline

\textbf{$U\ket{\psi_{\text{init}}}$}  & $-i\ket{0, 1}$ & $\ket{1, 1}$  & $\ket{0, -i~\alpha}$  &  $\ket{-i~\beta, -i~\alpha}$\\ \hline

\textbf{$U^{3/2}\ket{\psi_{\text{init}}}$} & -$\frac{1}{\sqrt{2}} (\ket{1, 0} + i \ket{0, 1})$ & $\frac{1}{\sqrt{2}} (\ket{2, 0} + \ket{0, 2})$ & $\ket{-\alpha/2, -i~\alpha/2}$ & $\ket{-\alpha/2 - i~\beta/2, -\beta/2 - i~\alpha/2}$ \\ \hline

\textbf{$U^2\ket{\psi_{\text{init}}}$}  & $-\ket{1, 0}$ & $\ket{1, 1}$ & $\ket{-\alpha, 0}$  & $\ket{-\alpha, -\beta}$   \\ \hline
\end{tabular}
\caption{\textbf{Cavity to cavity exchange}. For two cavities prepared to Fock states, the $U^p$ operator still behaves as a `beam splitter'. For the coherent state basis, it only exchanges coherent states between the two cavities, but will not create entanglement.}
\label{tab:cavTocav}
\end{table}

\newpage
\subsection{Multi-end router exchange}
One benefit of the router design is that we can apply multiple pump tones on the SNAIL at the same time to activate multiple exchange processes simultaneously. This give us enhanced freedom for more complicated gates. In order to test this idea, we have designed two experiments: one realizing multiple exchanges in parallel, and a second realizing exchange from one source cavity to two target cavities (V-exchange).

The protocol and experimental results are shown in Supplementary Figure \ref{fig:fancySwap}. All protocols are based on coherent states. First, we attempt the parallel exchange. Here, $C_3$ and $C_4$ have first been displaced with short Gaussian pulses, then the SNAIL pump is turned on such that $C_1-C_4$ and $C_2-C_3$ exchange is induced simultaneously. At the same time, the voltage output from all the cavities are measured through weakly-coupled ports. Comparing to the single exchange, a small, extra pump detuning ($\sim -250 \text{KHz}$) is observed on both pumps, which we attribute to the dynamic-Kerr shift from the driven SNAIL\cite{sivak_kerr-free_2019}.

For the ``V-exchange", we have displaced $C_2$ first, then we turn on the SNAIL pumps for $C_2-C_4$ and $C_2-C_3$ exchange simultaneously, and similarly, we have recorded the voltage output from all three cavities. One thing is different from parallel exchange is that the rate of two exchanges between each pair need to be precisely identical. In our case, as $C_3$ and $C_4$ have been perfectly overlapped with each other, thus, $C_2$ to $C_3$ and $C_2$ to $C_4$ share the same exchange rate with each other. Again, a small extra pump detuning ($\sim -60 \text{KHz}$) is added on both pumps to compensate for the dynamic-Kerr effect.

\begin{figure*}[h]
\includegraphics[]{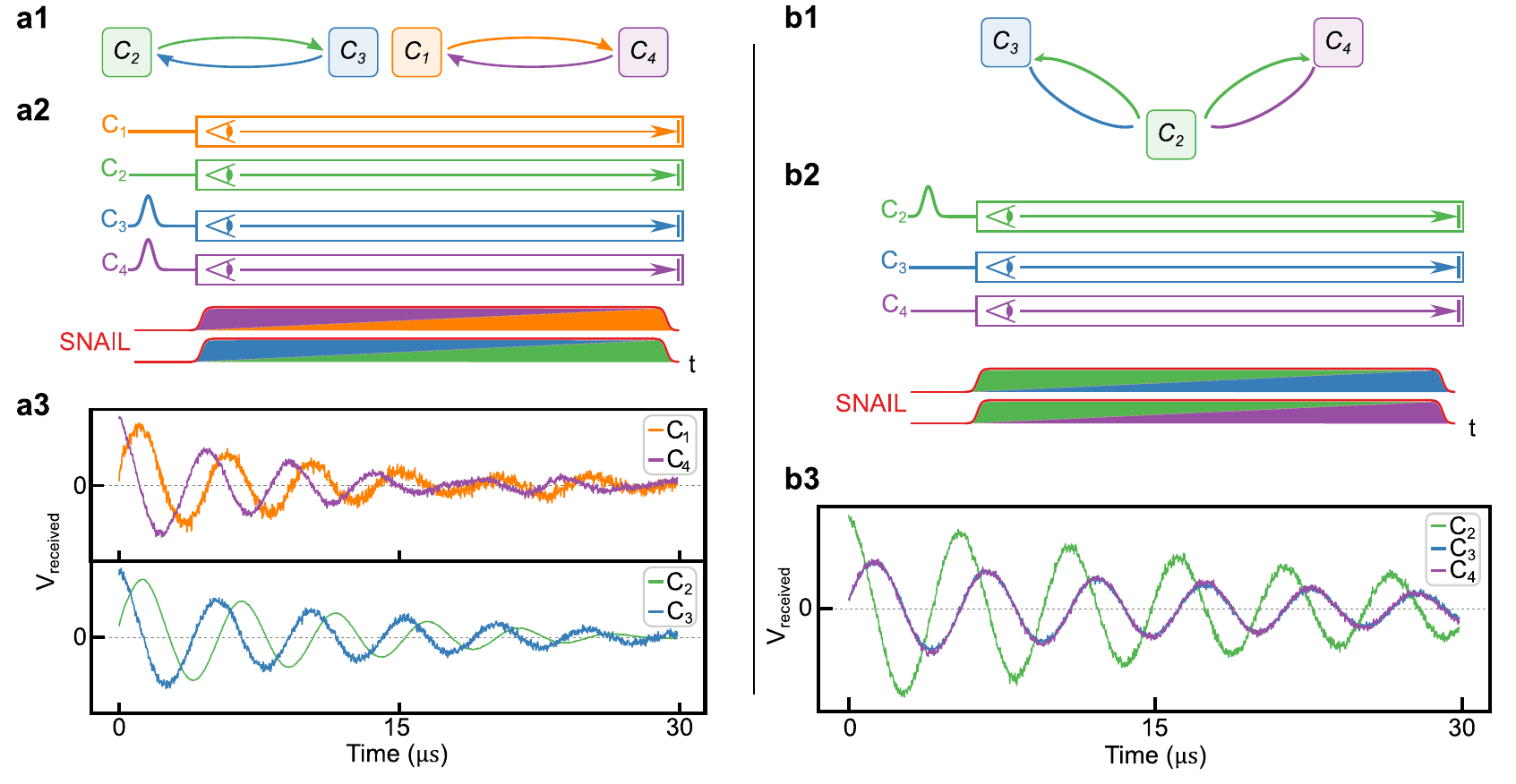}
\caption{\textbf{Simultaneous exchanges between multiple cavities.} (a) Parallel exchange between $C_1, C_4$ and $C_2, C_3$; (a1, a2) Pulse sequence and schematic. (a3) We have rotated the output results so that the quadrature components equal to zero, and plot only the in-phase voltages vs time for all four cavities. (b) V-exchange from $C_2$ to $C_3$ and $C_4$. In this case, the light can go back and forth between one cavity and two cavities. We have carefully tuned the pump frequencies and amplitudes so that the voltage trace for $C_3$ and $C_4$ perfectly overlap with each other. If $C_2$ is prepared to Fock state $\ket{1}$ at the beginning, and $C_3$ and $C_4$ are both prepared to empty states, then at the moment when $C_2$ is empty, a bell state would be created between $C_3$ and $C_4$, as we have shown in main text Sec. III}
\label{fig:fancySwap}
\end{figure*}
\clearpage

\newpage
\subsection{Inter-module GHZ state generation}
Other than the W-state preparation demonstrated in the main text, we also performed a GHZ state preparation experiment between all three modules' qubits. The protocal in shown in Supplementary Figure \ref{fig:ghz_state}a, this protocol requires one additional entangling gate: an intra-module CNOT. We achieve this using a state selective qubit $\pi$-pulse \cite{schuster2007resolving}.  We reconstruct the final state from tomography as shown in Supplementary Figure \ref{fig:ghz_state}b, and find a fidelity of $48.9 \pm 5.27~\%$, just below the threshold for provable entanglement. While the result falls below the threshold for provable entanglement, it points to our ability to implement an extensive gate set in the router, and create multi-qubit entanglement.

\begin{figure*}[h!]
\includegraphics[]{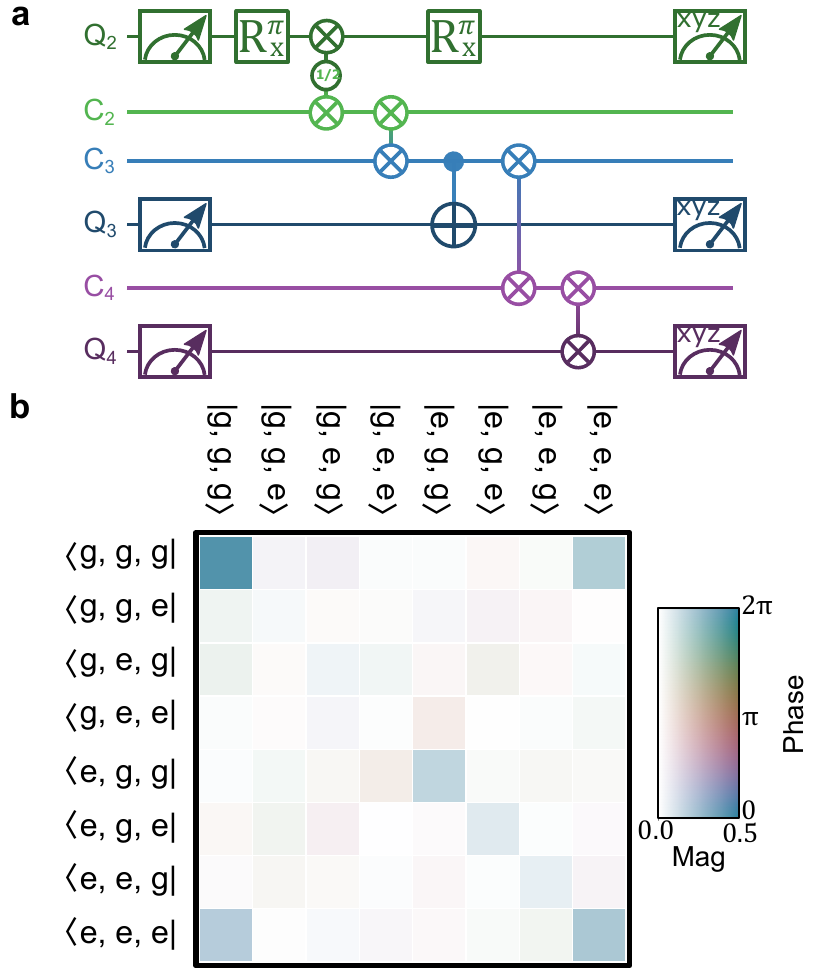}
\caption{\textbf{GHZ state generation experiment} (a) GHZ state generation sequence. First, $Q_2$ is prepared to excited state, then an intra\-module \rtsg ~ gate is performed between $Q_2$ and $C_2$, which creates a bell state between them. This entanglement is then transferred to $C_3$ via an inter-module $\sg$ gate. After that, an intra-module CNOT is performed between $C_3$ and $Q_3$, which entangles $Q_2$, $Q_3$ and $C_3$. In the end, $Q_2$ state is flipped with a $\pi$ pulse, and the photon in $C_3$ is transferred to $Q_4$ using two $\sg$ gate, which creates a GHZ state between $Q_2$, $Q_3$ and $Q_4$. (b) GHZ state generation density matrix reconstructed from tomography. Here, each element in the density matrix is represented with a color using the Hue-Chroma-Luminance (HCL) color scheme. The amplitude of each element is mapped linearly to the Chroma and Luminance of the color, and the phase (from 0 to 2$\pi$) is mapped linearly to the Hue value. This color mapping scheme has the property that elements of the same amplitude are perceived equally by the human eye, so that the small magnitudes fades into the white background to avoid drawing the eye to small, noisy matrix elements. The observed fidelity state is $48.9 \pm 5.27~\%$}
\label{fig:ghz_state}
\end{figure*}

\subsection{Generating Bell states during parallel operations}

Due to the coherence time of all modes, the Bell states' fidelity has been limited by the $\sg$ gate time in all operations. Here we use one example to show the limitation of the experiments. As we have mentioned in text, because of the decrease of saturation power when operating with multiple tones during parallel operations on the SNAIL, we need to slow down the $\sg$ rate. As shown in Supplementary Figure \ref{fig:fidelity}, when the $\sg$ gate time increase from $\unit[600]{ns}$ to $\unit[1300]{ns}$, the fidelity of bell states decrease from $76.9\%$ to $68.1\%$. However, no additional side effect is added due to cross-talking or other source.

\begin{figure*}[h]
\includegraphics[]{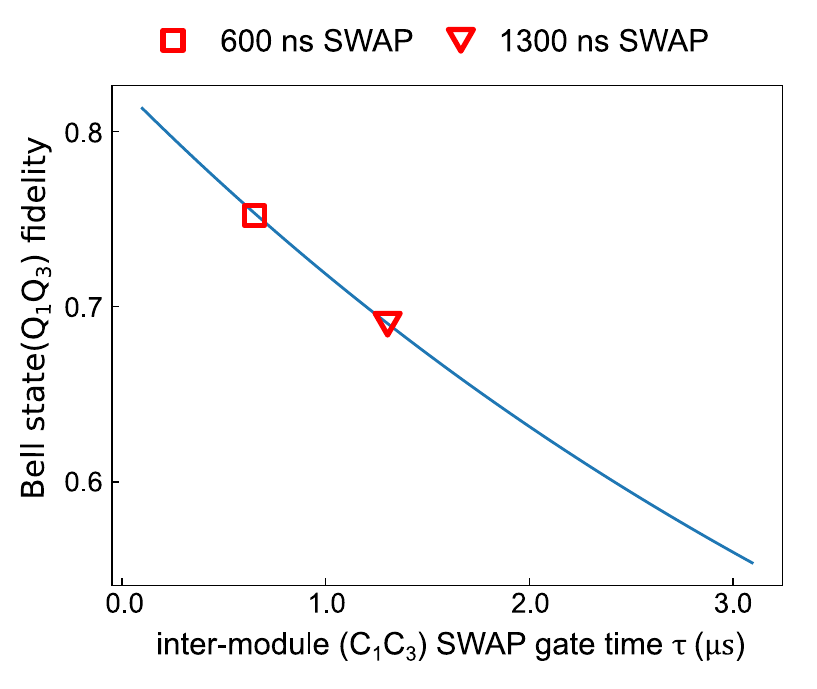}
\caption{\textbf{Bell stated fidelity versus inter-mdoule $\sg$ gate time.} The blue line represents the simulation of the bell state fidelity between $Q_1$ and $Q_3$ when we varying the inter-module $\sg$ gate time. The $\unit[600]{ns}$ ($\unit[1300]{ns}$) $\sg$ corresponding to the Bell state generation without (with) parallel operations.}
\label{fig:fidelity}
\end{figure*}

\newpage

\begin{center}
    \rule{10cm}{1pt}
\end{center}

\renewcommand\refname{Supplementary References}
\bibliography{refsSI}
\bibliographystyle{naturemag}